\documentclass[prl,twocolumn,aps,longbibliography,superscriptaddress,notoc]{revtex4-1}
\usepackage{bm}
\usepackage{graphicx}
\usepackage{amssymb}
\usepackage{amsmath}
\usepackage{eufrak}
\usepackage{color}
\usepackage[utf8]{inputenc}
\usepackage{ulem}
\usepackage[unicode=true,colorlinks=true,citecolor=blue,urlcolor=blue]{hyperref}

\renewcommand{\Re}{\mathop{\rm Re}}
\renewcommand{\Im}{\mathop{\rm Im}}

\newcommand{\e}{\mathrm{e}}

\let\ifr\i
\renewcommand{\i}{{\rm i}}
\renewcommand{\d}{\mathrm d}
\renewcommand{\emph}{\textit}
\newcommand{\braket}[1]{\left\langle #1 \right\rangle}

\usepackage{chngcntr}
\newcommand{\enquote}{}

\newcommand{\nix}[1]{}
\let\oldsec\section
\renewcommand{\section}[1]{\textit{#1}---}

\begin{document}

\title{Current Induced Hole Spin Polarization in Quantum Dot\\ via Chiral Quasi Bound State}

\author{V.~N.~Mantsevich}
\affiliation{Chair of Semiconductors and Cryoelectronics and Quantum Technology Center, Faculty of Physics, Lomonosov Moscow State University, 119991 Moscow, Russia}
\author{D.~S.~Smirnov}
\email[Electronic address: ]{smirnov@mail.ioffe.ru}
\affiliation{Ioffe Institute, Russian Academy of Sciences, 194021 St. Petersburg, Russia}

\begin{abstract}
We put forward a mechanism for current induced spin polarization for a hole in a quantum dot side-coupled to a quantum wire, that is based on the spin-orbit splitting of the valence band. We predict that in a stark contrast with the traditional mechanisms based on the linear in momentum spin-orbit coupling,
 an exponentially small bias applied to the quantum wire with heavy holes is enough to create the 100\% spin polarization of a localized light hole. Microscopically, the effect is related with the formation of chiral quasi bound states and the spin dependent tunneling of holes from the quantum wire to the quantum dot. This novel current induced spin polarization mechanism is equally relevant for the GaAs, Si and Ge based semiconductor nanostructures.
\end{abstract}

\maketitle{}

\section{Introduction}With the approach of the quantum computation era~\cite{dyakonov2020will} the localized spins in quantum dots (QDs) remain the most prominent candidates for the scalable quantum simulations and quantum information processing~\cite{Watson2018,Yoneda2018,Yang2020}. The electron spins can be already efficiently transferred in chains of QDs using the well controlled spin-spin interactions, entangled spin states can be generated, and CNOT-gates can be realized with very high fidelity~\cite{Nowack07,Zajac439,PhysRevLett.123.160501,Mills2019,PhysRevLett.126.017701,PhysRevLett.126.107401}.

However, electrical polarization of individual spins in QDs still remains a vital problem. Basically, this can be performed due to the pronounced spin-orbit interaction in most of semiconductors. Historically, the current induced spin polarization was first proposed theoretically and realized experimentally for bulk Te, which is a gyrotropic material~\cite{ivchenko1978new,vorob1979optical}. Later the current induced spin polarization was demonstrated for quantum wells made of GaAs-like semiconductors~\cite{Ganichev_110,silov04}, strained bulk semiconductors~\cite{PhysRevLett.93.176601}, and epilayers~\cite{Norman2014,doi:10.1063/1.4864468}. However, the maximum degree of current induced spin polarization is limited to a few percent because of the weakness of the momentum-dependent spin-orbit splitting compared to the Fermi energy~\cite{ganichev2012spin}. Streaming and hopping conductivity regimes can increase the polarization a few times~\cite{Golub2013,Hopping_spin}, but it still remains much smaller than unity.


In this Letter we demonstrate that the strong spin-orbit splitting of the valence band can be exploited to create 100\% spin polarization of holes localized in QDs in specifically designed structures. This splitting is large, for example, it is of the order of $300$~meV in GaAs and Ge and is about $70$~meV in Si. Thus, our proposal is equally relevant for high-quality optically-addressable GaAs-based structures~\cite{Huber2017,Chekhovich_register,Strong_coup_exp}, most technologically advanced Si-based structures~\cite{Veldhorst2015,Mi156,borjans2020resonant}, and emerging Ge-based structures~\cite{PhysRevB.84.195314,doi:10.1021/nl101181e,hendrickx2020fast,Ultrafast2021}.

Microscopically, the current induced spin polarization takes place due to the spin-dependent hole tunneling, which leads to the formation of chiral quasi bound states in continuum~\cite{Hsu2016,PhysRevLett.126.073001}. This concept was previously used for the spin filtering in magnetic field~\cite{VALLEJO20104928} and now is widely exploited in chiral photonics~\cite{Spitzer2018,Yin2020,Lin331}.





\begin{figure}
  \centering
  \includegraphics[width=0.95\linewidth]{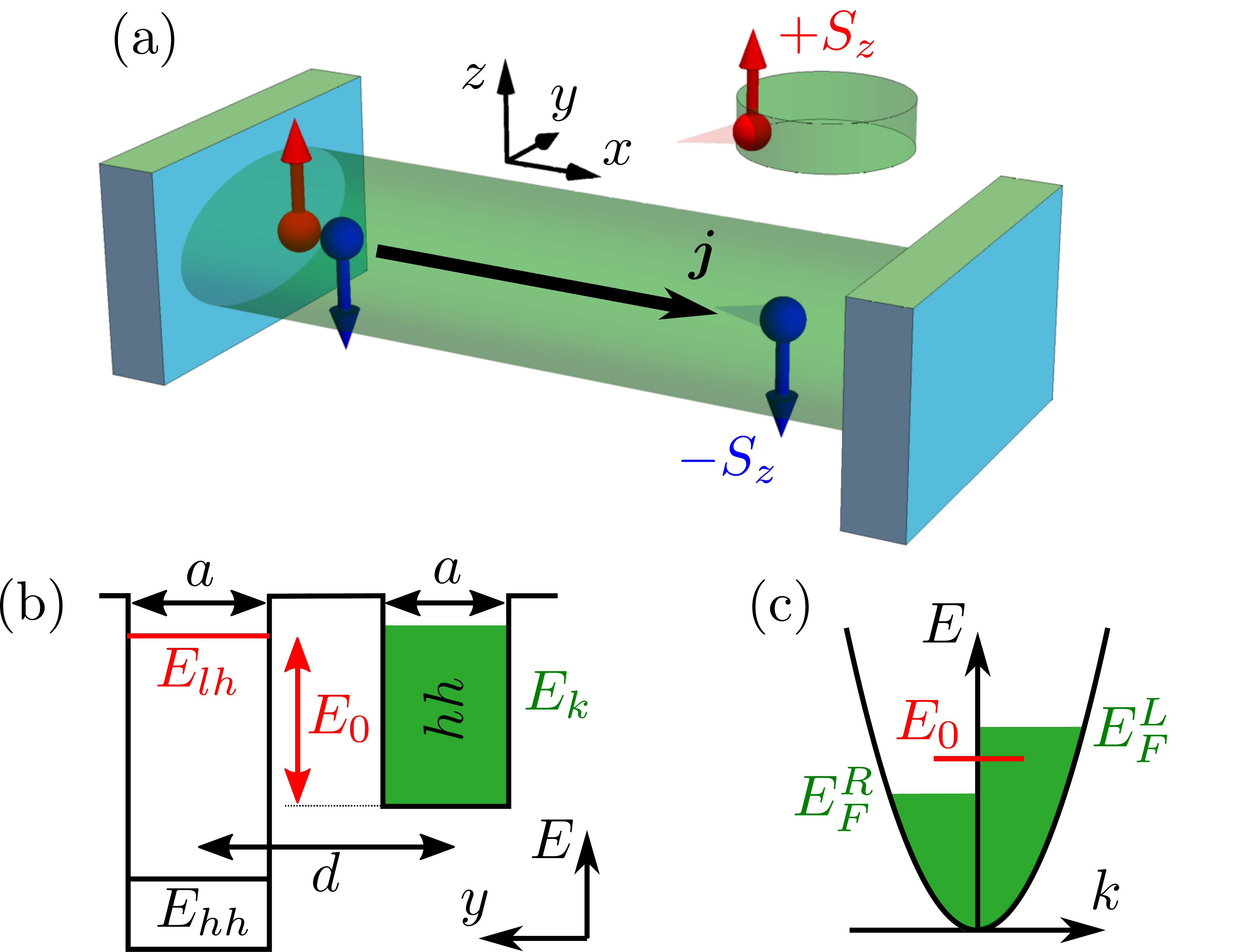}
  \caption{(a) QD side coupled to the quantum wire. The difference of the tunneling probabilities for spin-up (red balls with arrows) and spin-down (blue ones) holes from the quantum wire to the QD and of the spin flip rates leads to the current induced spin polarization in the QD. (b) Energy diagram and the geometric parameters of the system. 
    (c) Heavy hole distribution function in the quantum wire. The light hole resonance energy $E_0$ is between the Fermi energies of the heavy holes in the quantum wire propagating from the left, $E_F^L$, and from the right, $E_F^R$, leads.}
  \label{fig:system}
\end{figure}

\textit{System under study}
represents a quantum dot weakly side-coupled to a quantum wire~\cite{PhysRevLett.95.066801,Edlbauer2017,Borzenets2020}, see Fig.~\ref{fig:system}. We assume the structure to be formed electrostatically in a two-dimensional hole gas with a strong splitting between heavy and light hole subbands in the material with the top of the valence band described by $\Gamma_8$ representation of $T_d$ group or $\Gamma_8^+$ representation of $O_h$ group. We consider the Fermi energy of heavy holes in the wire to be close to the energy of light hole states localized in the QD, as shown in Fig.~\ref{fig:system}(b). The heavy hole state in the QD is assumed to be deeply below the Fermi energy, so this state is always doubly occupied. Thus we will take into account only the tunneling between heavy holes in the quantum wire and light holes in the QD.

The proposed device is described by the $C_{2v}$ symmetry group. We choose the coordinate frame to have the $x$ axis along the quantum wire and the $z$ axis along the structure growth axis, as shown in Fig.~\ref{fig:system}(a). Clearly, the electric current $j_x$ of heavy holes flowing along the wire can linearly couple to the light hole spin $S_z$ in the QD along the $z$ axis. The symmetry of this effect is the same as that of the Mott scattering or the spin Hall effect, so the spin polarization would change sign for the QD placed at the opposite side of the quantum wire. The current induced spin polarization does not require any magnetic field or microscopic symmetry reduction, i.e. Rashba or Dresselhaus spin-orbit interactions, and it appears even in the centrosymmetric materials such as Si and Ge along with GaAs. Because of this, a very large degree of spin polarization can be achieved, which we demonstrate below.


The Hamiltonian of the system can be written as follows:
\begin{multline}
  \label{eq:Ham}
  \mathcal H=E_0\sum_\pm n_\pm+Un_+n_-+\sum_{k,\pm}E_kn_{k,\pm}\\
  +\sum_{k,\pm}\left(V_{k,\pm}d_\pm^\dag c_{k,\mp}+{\rm H.c.}\right),
\end{multline}
where $E_0$ is a single light hole energy level in the QD, $n_\pm=d_\pm^\dag d_\pm$ are the occupancies of this state by holes with the spin $\pm1/2$ along the $z$ axis, respectively, with $d_\pm$ being the corresponding annihilation operators, $U$ is the Coulomb interaction energy between the two localized light holes, $E_k$ denotes the energy of a heavy hole in the quantum wire with the wave vector $k$, $n_{k,\pm}=c_{k,\pm}^\dag c_{k,\pm}$ are the occupancies of the states in the wire with the spin $\pm3/2$ with $c_{k,\pm}$ being the corresponding annihilation operators. We assume the wire to be ballistic and neglect the interaction between holes in it.

Most importantly, $V_{k,\pm}$ in Eq.~\eqref{eq:Ham} denote the tunneling matrix elements between the quantum wire and the QD. They are produced by the off-diagonal elements of the Luttinger Hamiltonian~\cite{PhysRev.102.1030,ivchenko05a} and can be calculated as follows:
\begin{equation}
  \label{eq:Vpm}
  V_{k,\pm}=-\frac{\sqrt{3}\gamma_2\hbar^2}{2m_0}\braket{\Phi_\pm|(k_x\mp\i k_y)^2|\Psi_{k,\mp}},
\end{equation}
where $\Phi_\pm$ denotes a localized light hole wave function of an isolated QD with the corresponding spin, $\Psi_{k,\pm}$ is a heavy hole wave function of an isolated quantum wire, $\gamma_2$ is the second Luttinger parameter (we use the spherical approximation), $m_0$ is the free electron mass, and $k_\alpha=-\i\partial/\partial\alpha$ with $\alpha=x,y$ are the components of the wave vector operator. It follows from the time reversal symmetry that $V_{k,\pm}=V_{-k,\mp}^*$.
We note that the heavy hole mass along the wire is given by $m=m_0/(\gamma_1+\gamma_2)$, where $\gamma_1$ is the first Luttinger parameter, and the dispersion is given by $E_k=\hbar^2k^2/(2m)$. We chose the state with $k=0$ to be the energy reference.

\begin{figure}
  \includegraphics[width=\linewidth]{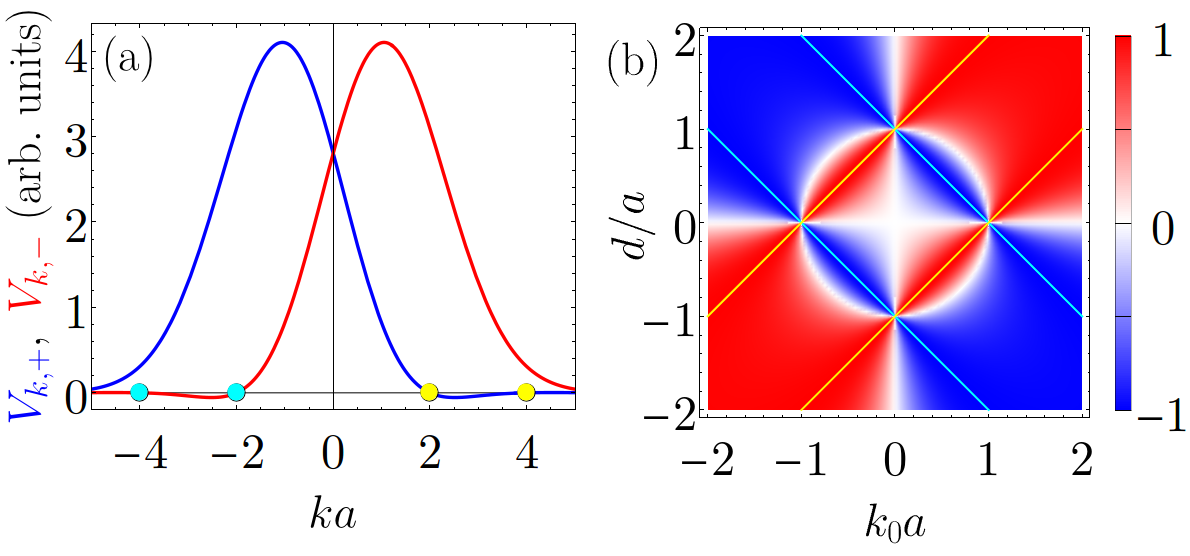}
  \caption{(a) Tunneling matrix elements calculated after Eq.~\eqref{eq:Vpm} for $d=3a$. (b) Chirality $\mathcal C$ of the quasi bound state calculated after Eq.~\eqref{eq:C}. The negative values of $d$ correspond to the location of the QD at the opposite side of the quantum wire. 
}
  \label{fig:chirality}
\end{figure}

To be specific, let us consider the Gaussian wave functions~\cite{supp}:
\begin{subequations}
  \label{eq:wave_functions}
  \begin{equation}
    \Phi_\pm\propto\exp\left(-\frac{x^2+(y-d)^2}{a^2}\right),
  \end{equation}
  \begin{equation}
    \Psi_{k,\pm}\propto\exp\left(\i kx-\frac{y^2}{a^2}\right),
  \end{equation}
\end{subequations}
where $a$ is the localization length, which is assumed to be the same for the QD and the quantum wire and $d$ is the distance between the center of the QD and the wire axis, see Fig.~\ref{fig:system}(b). Assumption of the same localization length for the QD and the quantum wire greatly simplifies the analytical calculations and does not change the general results. With these wave functions the matrix elements $V_{k,\pm}$ are real and can be calculated analytically~\cite{supp}. They are shown in Fig.~\ref{fig:chirality}(a) as functions of the wave vector for $d/a=3$. One can see that
for the given $k$ the tunneling matrix elements are generally strongly different for the spin-up and spin-down holes, which allows one to expect the high degree of the current induced spin polarization.



The single particle states of the Hamiltonian~\eqref{eq:Ham} are well known from the works of Anderson and Fano~\cite{PhysRev.124.41,fano61,Poddubny_Fano}. The coupling between the QD and the quantum wire leads to the formation of the quasi bound states at the energy $E_0$. Their chirality (also termed directionality) $\mathcal C$ can be defined as the difference of the probabilities for a light hole with a given spin to tunnel from the QD to the quantum wire states propagating to the right and to the left~\cite{lodahl2017chiral,Spitzer2018,PhysRevLett.126.073001}:
\begin{equation}
  \label{eq:C}
  \mathcal C=\frac{V_{k_0,+}^2-V_{-k_0,+}^2}{V_{k_0,+}^2+V_{-k_0,+}^2},
\end{equation}
where $k_0=\sqrt{2mE_0}/\hbar$. Due to the time reversal symmetry, the chirality is opposite for the two spin states, so the definition of its sign is ambiguous.

The chirality of the quasi bound state is shown in Fig.~\ref{fig:chirality}(b) as a color map. The regions with $k_0<0$ and $d<0$ show that the chirality is odd under reflection in $(yz)$ and $(xz)$ planes, respectively. Generally, the absolute value of the chirality is of the order of unity. Moreover, it turns to unity exactly along the four lines given by the equation
\begin{equation}
  \label{eq:C1}
  k_0a=\pm d/a\pm1,
\end{equation}
which are shown by yellow and light blue in Fig.~\ref{fig:chirality}(b). At the corresponding QD energies one of the tunneling matrix elements vanishes, which is shown by yellow and light blue circles in Fig.~\ref{fig:chirality}(a). As a result, completely chiral bound states in the continuum are formed, so that the light hole state in the QD couples to the states in the wire propagating only in one direction.
These completely chiral bound states in the continuum are robust and appear almost for any choice of the QD and the quantum wire wave functions, as we have checked.
The Gaussian form of the wave functions~\eqref{eq:wave_functions} corresponds to the parabolic localization potential, so at $k_0a>2$ the coupling to the second size quantized subband of the quantum wire can play a role. Still in the most realistic region of $k_0a<2$ and $d/a\gtrsim 2$ the chirality is very close to unity.

\section{Formalism}To calculate the current induced spin polarization in the nonequilibrium steady state, we use the nonequilibrium Keldysh diagram technique~\cite{RevModPhys.58.323,stefanucci_vanleeuwen_2013,Arseev_2015}. We start from the bare Hubbard retarded Green's function of an isolated QD~\cite{doi:10.1098/rspa.1963.0204,haug2008quantum}:
\begin{equation}
  \label{eq:G_R}
  G_{0,\sigma}^R(\omega)=\frac{1-\braket{n_{-\sigma}}}{\omega-E_0+\i\delta},
\end{equation}
where $\braket{n_{\sigma}}$ with $\sigma=\pm$ are the average occupancies of the corresponding light hole spin states, $\delta\to0^+$, and we measure frequencies in the units of energy for brevity. Henceforth we focus on the limit of strong Coulomb repulsion (as compared with the quasi bound state width), while the general case is described in the Supplemental Material~\cite{supp}. Then using the standard self energy $\Sigma_\sigma^R(\omega)=\sum_k|V_{k,\sigma}^2|/(\omega-E_k+\i\delta)$ we obtain from the Dyson equation
 \begin{equation}
   \label{eq:G_R_main}
   G_\sigma^R(\omega)=\frac{1-\braket{n_{-\sigma}}}{\omega-E_0+\i(1-\braket{n_{-\sigma}})\Gamma},
 \end{equation}
where $\Gamma=\pi D(E_0)(V_{k_0,+}^2+V_{k_0,-}^2)/4$ is the width of the quasi bound state with $D(E)=L\sqrt{2m/E}/(\pi\hbar)$ being the total density of states in the quantum wire. We assume $\Gamma$ to be much smaller than the band width $E_{1/a}$ and neglect the quasi bound state energy renormalization.

The occupancies of the QD states are given by the lesser Green's function:
\begin{equation}
  \label{eq:n_sigma_main}
  n_\sigma=-\i\int\frac{\d\omega}{2\pi}G_{\sigma}^<(\omega),
\end{equation}
which in the steady state is given by $G_{\sigma}^<=G_{\sigma}^R\Sigma_{\sigma}^<G_{\sigma}^A$. Here the lesser self energy depends on the Fermi energies in the left and right leads attached to the quantum wire, $E_F^L$ and $E_F^R$, respectively:
\begin{equation}
  \label{eq:Sigma_main}
  \Sigma_\sigma^<(\omega)=\frac{\pi\i D(\omega)}{2}\left[V_{k_0,\sigma}^2\theta(E_F^L-\omega)+V_{-k_0,\sigma}^2\theta(E_F^R-\omega)\right],
\end{equation}
where $\theta(\omega)$ is the Heaviside step function. Hence Eq.~\eqref{eq:n_sigma_main} yields
\begin{multline}
  \label{eq:set_main}
  \braket{n_\sigma}=(1-\braket{n_{-\sigma}})\left\{\frac{1+\sigma\mathcal C}{2\pi}\arctan\left[\frac{E_F^L-E_0}{(1-\braket{n_{-\sigma}})\Gamma}\right]
  \right.\\\left.
  +\frac{1-\sigma\mathcal C}{2\pi}\arctan\left[\frac{E_F^R-E_0}{(1-\braket{n_{-\sigma}})\Gamma}\right]+\frac{1}{2}\right\}.
\end{multline}
This set of two equations allows us to find self consistently the occupancies of the spin states $\braket{n_\sigma}$. Ultimately, the degree of the current induced spin polarization in the QD is given by $P=(\braket{n_+}-\braket{n_-})/(\braket{n_+}+\braket{n_-})$.

We note that this approach is valid on one hand for the temperatures below $\Gamma$, when the heavy holes distribution functions in the leads can be approximated by the step functions. On the other hand, the temperature is assumed to be larger than the Kondo temperature $T_K$, so that the high order correlations between holes can be neglected~\cite{haug2008quantum}. In fact the retarded Green's function~\eqref{eq:G_R_main} without quasi bound state width renormalization can be obtained from the equations of motion in the Hartree-Fock approximation~\cite{Lacroix_1981}. The Kondo effect at low temperatures can be taken into account, for example, using equations of motion truncated in an appropriate way at the temperatures of the order of $T_K$~\cite{PhysRevLett.70.2601,PhysRevB.68.195318} or much smaller than it~\cite{Lacroix_1981}. We expect that the Kondo effect would lead to the enhancement of the Coulomb blockade of one spin state by another, which would lead to the increase of the current induced spin polarization.

\begin{figure}
  \includegraphics[width=\linewidth]{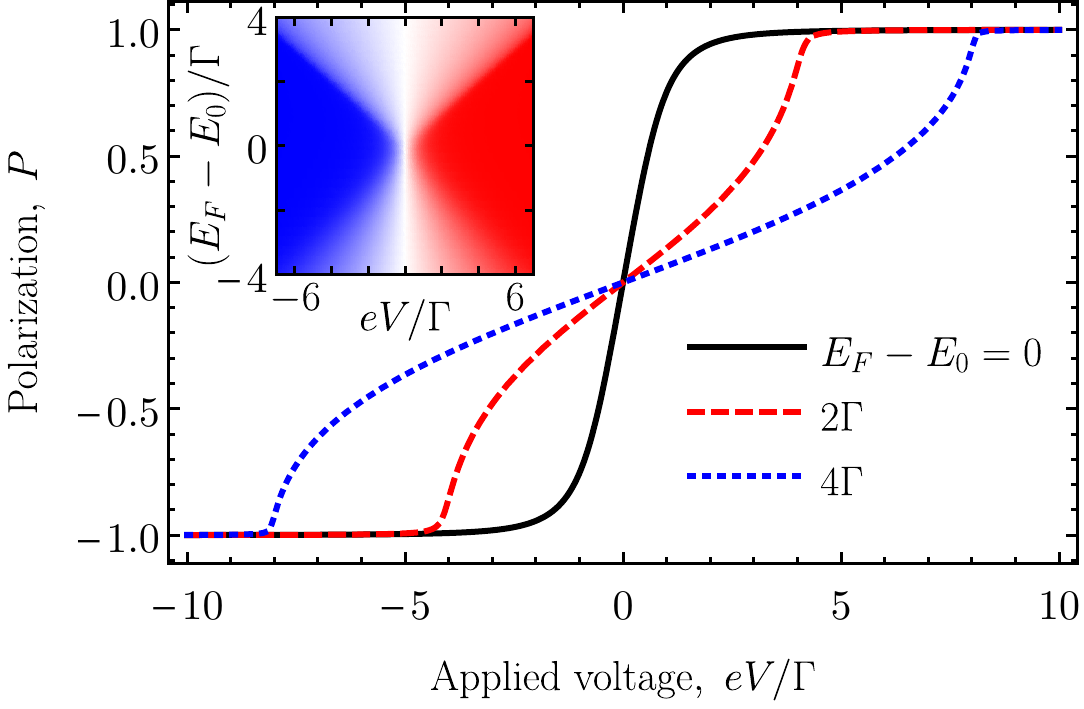}
  \caption{Current induced spin polarization in the chiral bound state, $\mathcal C=1$, as a function of the applied bias for the different Fermi levels $E_F-E_0$. The inset shows the spin polarization as a color map with the blue and red colors corresponding to $P=-1$ and $1$, respectively.}
  \label{fig:voltage}
\end{figure}

\section{Results}It follows from Eq.~\eqref{eq:set_main} that the chirality of the quasi bound state $\mathcal C$ produces explicit spin dependence of the occupancies provided a nonzero bias $eV=E_F^L-E_F^R$ is applied to the quantum wire. The current induced spin polarization is plotted in Fig.~\ref{fig:voltage} as a function of bias for the different positions of the Fermi energy $E_F=(E_F^L+E_F^R)/2$ for the completely chiral quasi bound state, $\mathcal C=1$. Generally, it is an odd function of the applied voltage and for the large voltages its absolute value reaches 100\%. For the resonant case, $E_F=E_0$, the spin polarization saturates at $eV\sim\Gamma$ and for the detuned case it saturates at $eV\sim2|E_F-E_0|$, as clearly seen from the inset in Fig.~\ref{fig:voltage}. Generally, the current induced spin polarization is the largest when the energy of the quasi bound state $E_0$ lies between the Fermi energies of the left and right leads, as it is shown in Fig.~\ref{fig:system}(c). For $\mathcal C<1$ the shape of this dependence is qualitatively the same.

Thus the large spin polarization can be induced by the voltages of the order of $\Gamma$, which is exponentially small being proportional to the squared tunneling matrix elements. The ultimate limit for it is set by the spin relaxation time in the isolated QD. In the absence of magnetic field, the nuclear spin fluctuations lead to the spin relaxation at the nanosecond time scale~\cite{book_Glazov,PRC}, which corresponds to $\Gamma\sim1~\mu$eV. However, to achieve this giant spin sensitivity the temperature of the system should be as low as a few millikelvins.

\begin{figure}[t]
  \includegraphics[width=\linewidth]{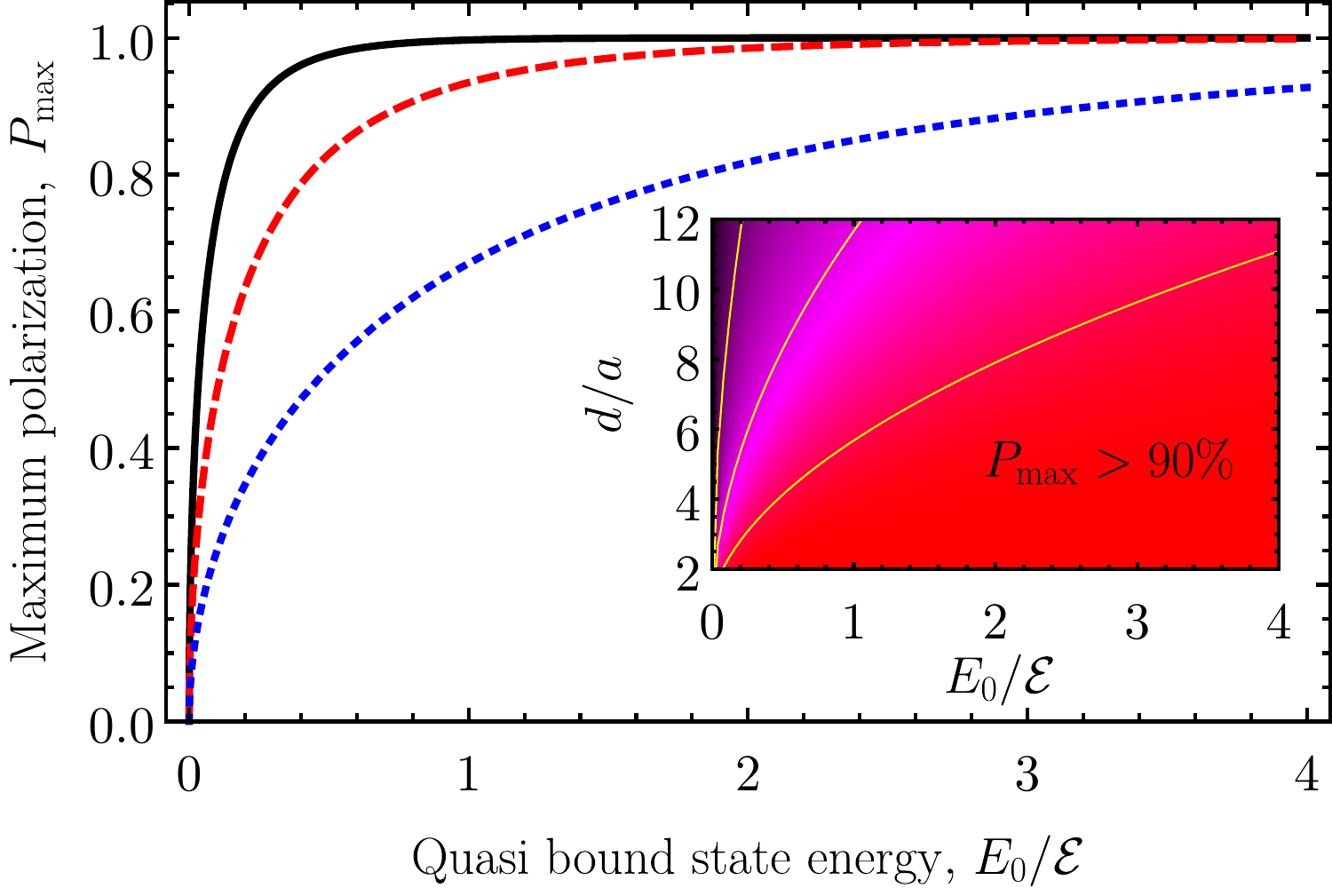}
  \caption{Spin polarization in the limit of the large bias as a function of the quasi bound state energy for $d/a=3$ (black solid curve), $5$ (red dashed curve), and $10$ (blue dotted curve). The inset shows the same as a color map. The yellow curves show the levels of $P_{\rm max}=0.3$, $0.6$ and $0.9$.}
  \label{fig:energy}
\end{figure}

The spin polarization in the limit of large bias can be found from Eq.~\eqref{eq:set_main} by setting the first arctangent to $\pi/2$ and the second one to $-\pi/2$. This gives
\begin{equation}
  \label{eq:P_max}
  P_{\rm max}=\frac{2\mathcal C}{\mathcal C^2+1},
\end{equation}
which is determined solely by the chirality of the quasi bound state. The maximum spin polarization degree is shown in Fig.~\ref{fig:energy} as a function of the quasi bound state energy for different distances between the QD and the quantum wire. Here the range of $E_0<4\mathcal E$ with $\mathcal E=\hbar^2/(2ma^2)$ corresponds to the energies below the bottom of the second size quantized band of the quantum wire. Generally, the maximum spin polarization decreases with increase of the distance between the quantum wire and the QD because of the chirality decrease shown in Fig.~\ref{fig:chirality}(b). However, it is still close to unity almost in the whole range of the quasi bound state energies $E_0$ up to $d\sim10a$. We note that this corresponds to the suppression of the quasi bound state width by a giant factor of $\e^{d^2/a^2}$. The inset in Fig.~\ref{fig:energy} explicitly shows the range of the parameters where the current induced spin polarization at the large bias exceeds 90\%.

\section{Discussion}The current induced spin polarization can be directly detected optically using the spin-induced Faraday rotation at the resonances related, for example, with the trion optical transitions in the QD. Alternatively, the spin polarization can be detected electrically using ferromagnetic contacts including magnetic microscope tips or using an additional QD weakly coupled to the main one due to the spin blockade effect and exchange interaction between the two QDs~\cite{hanson07}. There is also an inverse effect: the spin generation in the QD by external means would lead to the spin galvanic effect, e.g. to the current in the wire, which can be measured directly.

We note also that if the two QDs are placed at the opposite sides of the quantum wire, the electric current produces the opposite spin polarizations in them similarly to the spin Hall effect, as can be seen from the chirality shown in Fig.~\ref{fig:chirality}(b). However, for a single QD the hole spin flips are more efficient in one direction then in the opposite, so the total spin polarization of holes in the system is nonzero in contrast with the spin Hall effect.

We stress that the proposed mechanism of the current induced spin polarization via the chiral quasi bound states does not require magnetic field or linear in momentum spin-orbit coupling in contrast to the previous approaches. The spin-orbit interaction required here describes the splitting of the valence band exactly at the center of the Brillouin zone. It results in the off-diagonal matrix elements of the Luttinger Hamiltonain [see Eq.~\eqref{eq:Vpm}], which are related to the diagonal ones as $\gamma_2/\gamma_1$. We stress that this ratio is not parametrically small and amounts, for example, to $0.28$, $0.08$, and $0.32$ in GaAs, Si and Ge, respectively. The close to unity spin polarization is equally possible in all these materials.

\textit{In conclusion}, we demonstrated that the complex valence band structure leads to the chirality of the quasi bound states of the light holes in the QD side-coupled to the quantum wire with the heavy holes. The current flowing along the quantum wire produces nonequilibrium spin polarization of holes in the QD. This new mechanism of the current induced spin polarization has the following advantages: (i) It takes place for the exponentially small applied voltages. (ii) It does not require magnetic field and can be applied to the centrosymmetric materials. (iii) In the broad range of the geometric parameters the current induced spin polarization reaches 100\%.

We thank \href{https://orcid.org/0000-0003-3818-1014}{L. E. Golub}, \href{https://orcid.org/0000-0003-4462-0749}{M. M. Glazov}, and \href{http://www.ioffe.ru/theory/sector/krainov.html}{I. V. Krainov} for fruitful discussions. We acknoledge RF President Grant No. MK-5158.2021.1.2 and the Foundation for the Advancement of Theoretical Physics and Mathematics ``BASIS''. Analytical calculations by D.S.S were supported by the Russian Science Foundation grant No. 19-12-00051. V.N.M. acknowledges support by RFBR grant ($20-32-70001$ Stability) and support from the Interdisciplinary Scientific and Educational School of Moscow University ``Photonic and Quantum technologies. Digital medicine''.


\renewcommand{\i}{\ifr}
\let\oldaddcontentsline\addcontentsline
\renewcommand{\addcontentsline}[3]{}

%

\let\addcontentsline\oldaddcontentsline
\makeatletter
\renewcommand\tableofcontents{%
    \@starttoc{toc}%
}
\makeatother
\renewcommand{\i}{{\rm i}}

\appendix

\onecolumngrid
\vspace{\columnsep}
\begin{center}
\newpage
\makeatletter
{\large\bf{Supplemental Material to\\``\@title''}}
\makeatother
\end{center}
\vspace{\columnsep}

The Supplementary Material includes the following topics:

\hypersetup{linktoc=page}
\tableofcontents
\vspace{\columnsep}
\twocolumngrid

\counterwithin{figure}{section}
\renewcommand{\section}[1]{\oldsec{#1}}
\renewcommand{\thepage}{S\arabic{page}}
\renewcommand{\theequation}{S\arabic{equation}}
\renewcommand{\thefigure}{S\arabic{figure}}

\setcounter{page}{1}
\setcounter{section}{0}
\setcounter{equation}{0}
\setcounter{figure}{0}

\section{S1. Hamiltonian and matrix elements}

The Hamiltonian of the quantum dot (QD) with light holes side-coupled to the quantum wire with heavy holes can be written as follows [Eq.~\eqref{eq:Ham} in the main text]:
\begin{multline}
  \label{eq:Ham_S}
  \mathcal H=E_0\sum_\pm n_\pm+Un_+n_-+\sum_{k,\pm}E_kn_{k,\pm}\\
  +\sum_{k,\pm}\left(V_{k,\pm}d_\pm^\dag c_{k,\mp}+{\rm H.c.}\right).
\end{multline}
We recall that $n_\pm=d_\pm^\dag d_\pm$ are the occupancies of the light holes states in the QD having the spins $J_z=\pm1/2$ with $d_\pm$ ($d_\pm^\dag$) being the corresponding annihilation (creation) operators, $E_0$ is the energy of the light hole states in the QD which includes the interaction with the lower lying occupied heavy hole states in the QD, $U$ is the Coulomb repulsion energy between the two light hole states, $n_{k,\pm}=c_{k,\pm}^\dag c_{k,\pm}$ are the occupancies of the heavy hole states in the quantum wire with the wave vector $k$ and spin $J_z=\pm3/2$ with $c_{k,\pm}$ ($c_{k,\pm}^\dag$) being the corresponding annihilation (creation) operators, and, finally, $V_{k,\pm}$ are the tunneling matrix elements.

The tunneling matrix elements $V_{k,\pm}$ can be calculated using the Luttinger Hamiltonian, which in the hole representation has the form~\cite{ivchenko05a}:
\begin{equation}\label{LuttingerHamiltonian}
  \mathcal H_L = \left(
    \begin{array}{cccc}
        F & H & I & 0\\
        H^*& G & 0 & I\\
        I^*& 0 & G & -H\\
        0 & I^*& -H^*& F
    \end{array}
    \right).
\end{equation}
It is written in the basis of the spin states $J_z=+3/2,+1/2,-1/2,-3/2$ and has the elements
\begin{eqnarray}
    F &=&\frac{\hbar^2(\gamma_1 - 2\gamma_2)}{2m_0} k_z^2
    + \frac{\hbar^2(\gamma_1 + \gamma_2)}{2m_0}(k_x^2 + k_y^2),\nonumber\\
    G &=&\frac{\hbar^2(\gamma_1 + 2\gamma_2)}{2m_0} k_z^2
    + \frac{\hbar^2(\gamma_1 - \gamma_2)}{2m_0}(k_x^2 + k_y^2),\nonumber\\
    H &=& -\frac{\sqrt{3}\hbar^2\gamma_2}{m_0} k_z (k_x - i k_y),\nonumber\\
    I &=& -\frac{\sqrt{3}\hbar^2}{2m_0}\left[\gamma_2(k_x^2 - k_y^2) -2i\gamma_3k_xk_y
    \right],
\end{eqnarray}
where $m_0$ is the free electron mass, $\gamma_{1,2,3}$ are the Luttinger parameters and $\bm k$ is the hole wave vector. We use the spherical approximation $\gamma_2=\gamma_3$, so the Hamiltonian takes the form
\begin{equation}
  \mathcal H_L=\frac{\hbar^2k^2}{2m_0}\left(\gamma_1+\frac{5}{2}\gamma_2\right)-\frac{\hbar^2\gamma_2}{m_0}(\bm{k J})^2,
\end{equation}
where $\bm J$ is the hole spin. In this case the energy of the heavy hole states in the quantum wire reads
\begin{equation}
  E_k=\frac{\hbar^2k^2}{2m}
\end{equation}
with $m=m_0/(\gamma_1+\gamma_2)$ being the heavy hole mass along the wire.

Let $\hat\Psi_{k,\pm}=\Psi_{k,\pm}\chi_{\pm3/2}$ be the heavy hole wave function in the quantum wire with the wave vector $k$ along the wire and the spin $J_z=\pm3/2$ and $\hat\Phi_{\pm}=\Phi_\pm\chi_{\pm1/2}$ be the light hole wave function in the QD with the spin $J_z=\pm1/2$, respectively, where $\chi_{J_z}$ are the spinors. The tunneling matrix elements between them involve the change of the spin, which can not be provided by the external electrostatic potential. Instead, it is produced by the Luttinger Hamiltonian:
\begin{equation}
  V_{k_\pm}=\braket{\hat\Phi_{\pm}|\mathcal H_L|\hat\Psi_{k,\mp}}.
\end{equation}
We assume that structure is symmetric in the $(xy)$ plane containing the QD and the quantum wire, so the matrix elements between the states $\hat\Phi_{\pm}$ and $\hat\Psi_{k,\pm}$, respectively, being given by the $H$ term in the Luttinger Hamiltonian, which is proportional to $k_z$, vanish. As a result, the coupling takes place between the states $\hat\Phi_\pm$ and $\hat\Psi_{k,\mp}$ only. It is produced by the matrix element $I$ and involves the spin flip from $\mp3/2$ to $\pm1/2$, respectively.

\begin{figure}[t]
  \centering
  \includegraphics[width=0.5\linewidth]{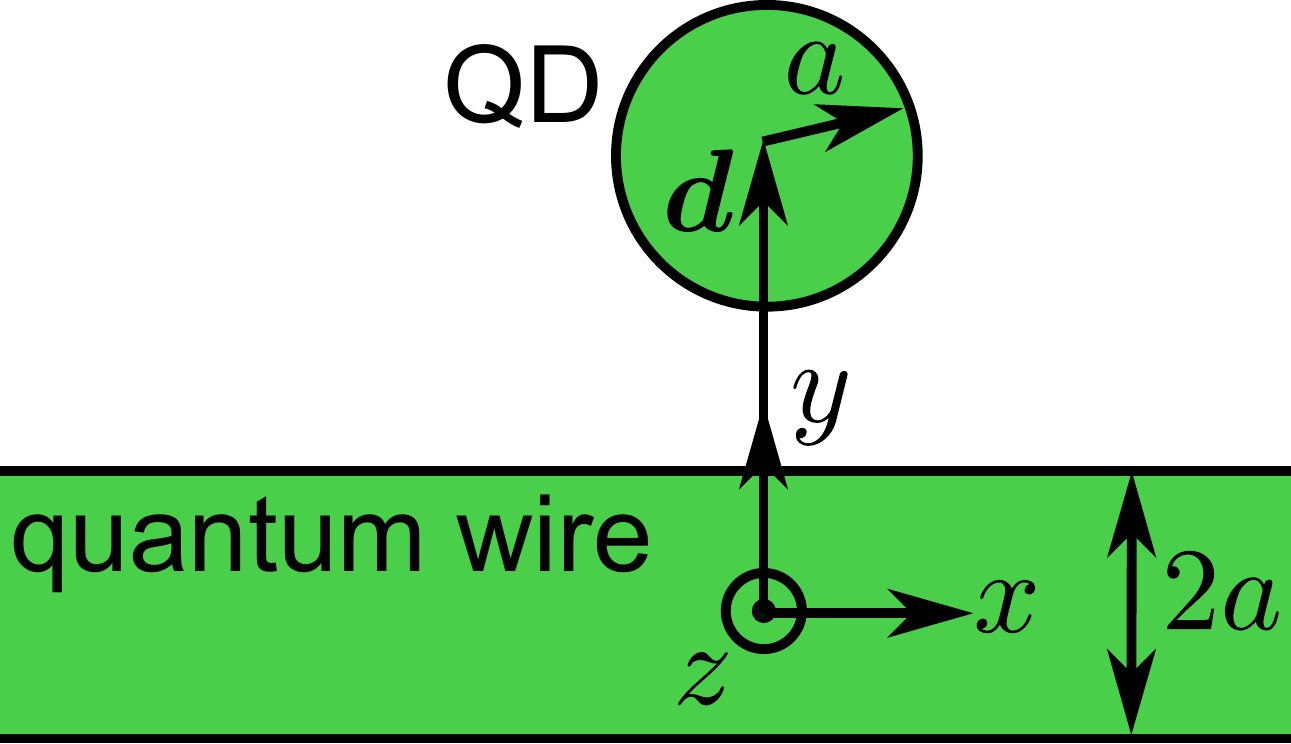}
  \caption{Geometry of the system and the coordinate frame.}
  \label{fig:coordinates}
\end{figure}

To be specific, we consider the Gaussian wave functions [see Eqs.~\eqref{eq:wave_functions} in the main text]
\begin{subequations}
  \label{eq:Psi_Gaussian}
  \begin{equation}
    \Phi_\pm=\varphi(z)\sqrt{\frac{2}{\pi}}\frac{1}{a}\exp\left(-\frac{x^2+(y-d)^2}{a^{2}}\right),
  \end{equation}
  \begin{equation}
    \label{eq:Psi}
    \Psi_{k,\pm}=-\psi(z)\sqrt{\frac{1}{aL}}\sqrt[4]{\frac{2}{\pi}}\exp\left(\i k x-\frac{y^{2}}{a^{2}}\right).
  \end{equation}
\end{subequations}
Here we use the coordinate frame with the origin at the center of the quantum wire cross section, we choose the $x$ axis to be parallel to the quantum wire and the QD center to be located at the coordinates $(0,d,0)$, see Fig.~\ref{fig:coordinates}. We assume the localization length $a$ to be the same for the QD and the quantum wire, $L$ is the normalization length, $\varphi(z)$ and $\psi(z)$ are the normalized wave functions along the growth axis $z$, and we assume the size quantization in this direction to be the strongest. The minus sign is introduced in Eq.~\eqref{eq:Psi} in order to get mostly positive tunneling matrix elements. We note that the matrix elements for the different localization lengths of the QD and the quantum wire can be also calculated analytically. However, we do not demonstrate it here since the corresponding expressions are cumbersome and all the physical effects do not change qualitatively.

For the wave functions~\eqref{eq:Psi_Gaussian} the tunneling matrix elements read
\begin{multline}
  \label{eq:V_pm}
  V_{k,\pm}=\frac{\sqrt{3}\sqrt[4]{2\pi}\gamma_2\hbar^2V_z}{2m_0a\sqrt{La}}\left[\left(\frac{d}{a}\pm ka\right)^2-1\right]\\
  \times\exp\left(-\frac{a^{2}k^{2}}{4}-\frac{d^{2}}{2a^{2}}\right),
\end{multline}
where $V_z=\braket{\phi(z)|\psi(z)}$. One can see, that the matrix elements are real. The matrix elements exponentially decay with the distance $d$ between the QD and the quantum wire. Most importantly they are different for the given $k$, as illustrated in Fig.~\ref{fig:chirality}(a) in the main text. We note that the wave function~\eqref{eq:Psi} corresponds to the harmonic localization potential across the quantum wire along the $y$ direction. In this case the energy of the state with the wave vector $k=2/a$ coincides with the bottom of the second size quantized subband, so we will consider below the states in the range of $ka<2$ only. In the same time, the ratio $d/a$ can be arbitrary large in the model, but in fact it should not be too large in order to observe the effects of the hole tunneling between the QD and the quantum wire. We stress, that the tunneling matrix elements $V_{k,\pm}$ involve hole spin flips, but contain $\gamma_2/\gamma_1$ as a factor describing the spin-orbit interaction only. This ratio is not small, so the current induced spin polarization in this system is not parametrically suppressed.

For the given energy of the light hole states in the QD $E_0$ and the corresponding wave vector $k_0=\sqrt{2mE_0}/\hbar$ the tunneling matrix elements are different $V_{k_0,+}\neq V_{k_0,-}$, which allows us to define chirality as follows [Eq.~\eqref{eq:C} in the main text]~\cite{lodahl2017chiral,Spitzer2018,PhysRevLett.126.073001}:
\begin{equation}
  \label{eq:C_S}
  \mathcal C=\frac{V_{k_0,+}^2-V_{k_0,-}^2}{V_{k_0,+}^2+V_{k_0,-}^2}.
\end{equation}
We note that due to the mirror and time reversal symmetries
\begin{equation}
  \label{eq:symm}
  V_{k,\pm}=V_{-k,\mp},
\end{equation}
so the definition of the sign of $\mathcal C$ is arbitrary.

From Eq.~\eqref{eq:V_pm} we find the chirality
\begin{equation}
  \mathcal C=\frac{4dk_0[(k_0a)^2+(d/a)^2-1]}{[(k_0a)^2-1]^2+6(dk_0)^2-2(d/a)^2+(d/a)^4}.
\end{equation}
One can see that it is an odd function of $d$ and $k_0$ as shown in Fig.~\ref{fig:chirality}(b) in the main text. It vanishes at $d=0$ and $k=0$, as expected, and also at $(d/a)^2+(ka)^2=1$. These lines are white in Fig.~\ref{fig:chirality}(b) in the main text. But most importantly the chirality reaches $\pm1$ at [Eq.~\eqref{eq:C1} in the main text]
\begin{equation}
  k_0a=\pm d/a\pm1,
\end{equation}
which is shown by yellow and light blue lines in Fig.~\ref{fig:chirality}(b) in the main text. Along these lines the current induced spin polarization can reach exactly $100\%$, as we demonstrate in the main text. We note that the corresponding energy $E_0$ can be found for any $d$, and we checked that this holds for a broad class of the wave functions apart from Eqs.~\eqref{eq:Psi_Gaussian}.

\section{S2. General formalism for current induced spin polarization}

Here we calculate the current induced spin polarization in the QD produced by the nonequilibrium distribution functions of the heavy holes in the quantum wire inherited from the attached leads. We assume that the occupancies of the states in the quantum wire are given by
\begin{equation}
  \label{eq:nk}
  \braket{n_k}=\theta(E_k)\left[\theta(E_F^L-E_k)\theta(k)+\theta(E_F^R-E_k)\theta(-k)\right],
\end{equation}
where $E_F^L$ and $E_F^R$ are the Fermi energies in the left and right leads, respectively,  and $\theta(t)$ is the Heaviside step function. This distribution is relevant for the temperatures below the width of the quasi bound state.

At the first step, we neglect tunneling and use equations of motion to obtain the retarded Hubbard Green's function of the QD~\cite{doi:10.1098/rspa.1963.0204,haug2008quantum} [c.f. Eq.~\eqref{eq:G_R} in the main text]: 
\begin{equation}
  \label{eq:Hubbard}
  G_{0,\sigma}^R(\omega)=\frac{1-\braket{n_{-\sigma}}}{\omega-E_0+\i\delta}+\frac{\braket{n_{-\sigma}}}{\omega-E_0-U+\i\delta},
\end{equation}
where we set $\hbar=1$ for brevity. We recall that it is defined as the Fourier transform of $-\i\braket{\left\{d_\sigma^\dag(0),d_\sigma(t)\right\}}\theta(t)$. The average occupancies of the light hole spin states $\braket{n_\sigma}$ should be determined self consistently taking into account the tunneling processes.

\begin{figure}[t]
  \centering
  \includegraphics[width=\linewidth]{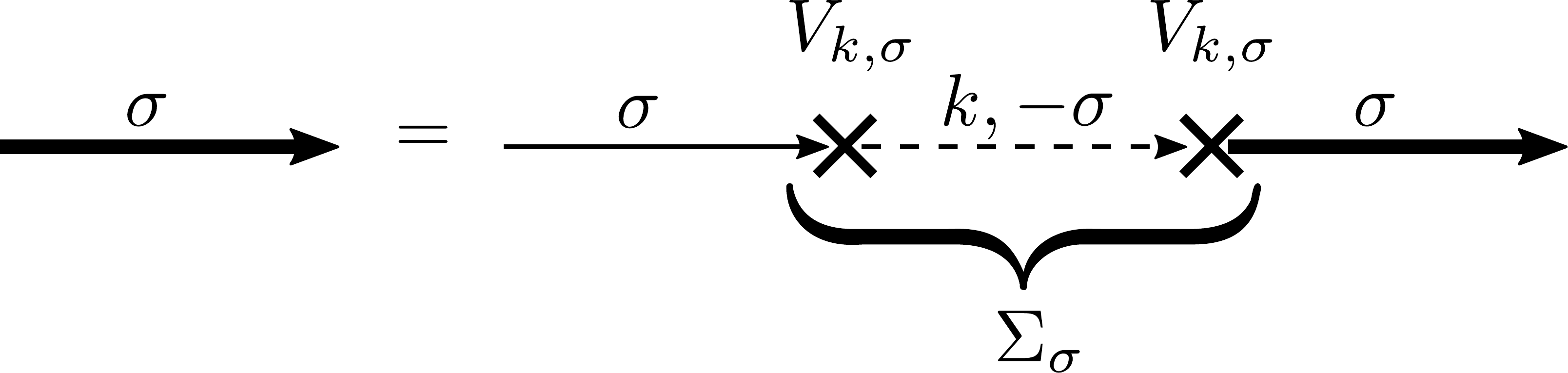}
  \caption{Dyson equation and the self energy for the tunneling problem. Thick solid, thin solid and dashed lines correspond to $G_\sigma(\omega)$, $G_{0,\sigma}(\omega)$, and $G_{k,-\sigma}(\omega)$, respectively.}
  \label{fig:self}
\end{figure}

Then we account for the tunneling as a perturbation while keeping the Fermi energies in the left and right leads, $E_F^{L}$ and $E_F^R$, different. This can be done in the Keldysh formalism~\cite{RevModPhys.58.323,stefanucci_vanleeuwen_2013,Arseev_2015}. The self energy in this problem is trivial, see Fig.~\ref{fig:self}, it reads
\begin{equation}
  \label{eq:Sigma_R}
  \Sigma_\sigma^R(\omega)=\sum_{k}|V_{k,\sigma}^2|G_{k,-\sigma}^R(\omega),
\end{equation}
where
\begin{equation}
  G_{k,\sigma}^R(\omega)=\frac{1}{\omega-E_k+\i\delta}
\end{equation}
is the retarded Green's function of heavy holes in the quantum wire. We note that the tunneling in Eq.~\eqref{eq:Ham_S} flips the spin $\sigma$. From the Dyson equation, Fig.~\ref{fig:self}, we obtain
\begin{equation}
  \label{eq:GR}
  G_{\sigma}^R(\omega)=\frac{G_{0,\sigma}^R(\omega)}{1-G_{0,\sigma}^R(\omega)\Sigma_\sigma^R(\omega)}.
\end{equation}

We note that the retarded self energy $\Sigma_{\sigma}^R(\omega)$ in Eq.~\eqref{eq:Sigma_R} diverges at $\omega=-\i\delta$ and as a result one obtains a pole in $G_{\sigma}^R(\omega)$ at small negative real $\omega$, which corresponds to the truly bound state. This state is generally present in the one dimensional problems with an impurity. We will assume that $E_0$ is large enough and will neglect the contribution of the true bound state to the current induced spin polarization. In this case, in the steady state the lesser Green's function defined as the Fourier transform of $\i\braket{d_\sigma^\dag d_\sigma(t)}$ is given by
\begin{equation}
  \label{eq:Gm}
  G_{\sigma}^<(\omega)=G_{\sigma}^R(\omega)\Sigma_{\sigma}^<(\omega)G_{\sigma}^A(\omega),
\end{equation}
where $G_\sigma^A(\omega)=[G_\sigma^R(\omega)]^*$ is the advanced Green's function and $\Sigma_{\sigma}^<(\omega)$ is the lesser self energy given by
\begin{equation}
  \Sigma_\sigma^<(\omega)=\sum_{k}|V_{k,\sigma}^2|G_{k,-\sigma}^<(\omega),
\end{equation}
as follows from Fig.~\ref{fig:self}. The lesser Green's functions of the holes in the quantum wire are determined by the occupancies of the states:
\begin{equation}
  G_{k,\sigma}^<(\omega)=2\pi\i\braket{n_k}\delta(\omega-E_k),
\end{equation}
which are defined by the Fermi energies in the leads, see Eq.~\eqref{eq:nk}. This gives the lesser self energy
\begin{equation}
  \label{eq:Sigma_m}
  \Sigma_\sigma^<(\omega)=2\i\left[\Gamma_{L,\sigma}(\omega)n_L(\omega)+\Gamma_{R,\sigma}(\omega)n_R(\omega)\right],
\end{equation}
where
\begin{equation}
  \Gamma_{L/R,\sigma}(\omega)=\pi\frac{D(\omega)}{4}V_{+k_\omega/-k_\omega,\sigma}^2
\end{equation}
are the tunneling rates with $D(\omega)=(L/\pi)\sqrt{2m/\omega}$ being the total density of states in the quantum wire including spin, $n_{L/R}(\omega)=\braket{n_{+k_\omega/-k_\omega}}$, and $k_\omega=\sqrt{2m\omega}$. We note that $\Gamma(\omega)=\Gamma_{L,\sigma}(\omega)+\Gamma_{R,\sigma}(\omega)$ determines the widths of the quasi bound states according to:
\begin{equation}
  -\Im\left[\Sigma_\sigma^R(\omega)\right]=\Gamma(\omega),
\end{equation}
where we used Eq.~\eqref{eq:Sigma_R}. It does not depend on spin, as follows from Eq.~\eqref{eq:symm}. The tunneling rates $\Gamma_{L/R,\sigma}(\omega)$ can be also obtained from the Fermi golden rule.

Finally, the occupancies of the spin states in the QD $\braket{n_\pm}$ should be found self consistently as
\begin{equation}
  \label{eq:n_pm}
  \braket{n_\sigma}=-\i\int\limits_{-\infty}^\infty G_\sigma^<(\omega)\frac{\d\omega}{2\pi},
\end{equation}
where the lesser Green's function is given by Eq.~\eqref{eq:Gm} with the lesser self energy from Eq.~\eqref{eq:Sigma_m} and retarded and advanced Green's functions from Eq.~\eqref{eq:GR}, which includes the same occupancies of the spin states through the Hubbard Green's function, Eq.~\eqref{eq:Hubbard}.

Ultimately, the spin polarization in the QD is given by
\begin{equation}
  \label{eq:P}
  P=\frac{\braket{n_+}-\braket{n_-}}{\braket{n_+}+\braket{n_-}}.
\end{equation}
Below we consider the limits of zero and infinite Coulomb interaction and use the wide band approximation to obtain simplified expressions for the current induced spin polarization.

\section{S3. Weak Coulomb interaction}

Here for the purpose of illustration we consider the simple limit of $U=0$, when the interaction between the holes in the QD can be neglected. In this limit the bare Green's function of the QD, Eq.~\eqref{eq:Hubbard}, reduces to
\begin{equation}
  G_{0,\sigma}^R(\omega)=\frac{1}{\omega-E_0+\i\delta}.
\end{equation}
Thus from Eq.~\eqref{eq:GR} we obtain
\begin{equation}
  \label{eq:G_R_U0}
  G_{\sigma}^R(\omega)=\frac{1}{\omega-\tilde E_0(\omega)+\i\Gamma(\omega)},
\end{equation}
where
\begin{equation}
  \tilde E_0(\omega)=E_0+\Re\left[\Sigma_\sigma^R(\omega)\right]
\end{equation}
is the energy of the quasi bound state in the QD. Substituting it in Eq.~\eqref{eq:Gm} and Eq.~\eqref{eq:n_pm} along with Eq.~\eqref{eq:Sigma_m} we obtain the occupancies of the spin states
\begin{equation}
  \label{eq:n_pm_0}
  \braket{n_\pm}=\int\limits_{0}^\infty\frac{2\Gamma(\omega)n_\pm(\omega)}{\left[\omega-\tilde E_0(\omega)\right]^2+\Gamma^2(\omega)}\frac{\d\omega}{2\pi},
\end{equation}
where
\begin{equation}
  n_\pm(\omega)=\frac{\Gamma_{L,\pm}(\omega)n_L(\omega)+\Gamma_{R,\pm}(\omega)n_R(\omega)}{\Gamma(\omega)}.
\end{equation}

One can obtain the same result for the limit of negligible Coulomb interaction from the exact solution of a single particle problem. To explicitly demonstrate this, we consider the single particle eigenfunctions of the Hamiltonian~\eqref{eq:Ham_S} with $U=0$. They can be found following, for example, the original work of Fano~[\onlinecite{fano61}]. For the given energy $E>0$ and spin $\sigma$ of the light hole state, the two energy degenerate wave functions have the form
\begin{multline}
  \label{eq:Psis}
  \Psi_{1,2}^\sigma=a_{1,2}\Phi_{\sigma}+{\rm v.p.}\int\limits_0^\infty\d E'\left[b_{1,2}^\sigma(E')\Psi_{k',-\sigma}\right.\\\left.+c_{1,2}^\sigma(E')\Psi_{-k',-\sigma}\right],
\end{multline}
where the coefficients are
\begin{subequations}
  \begin{equation}
    a_1=\frac{4\sin(\Delta)}{\pi D(E)Y_E},
  \end{equation}
  \begin{equation}
    b_1^\sigma(E')=\frac{V_{k',\sigma}}{Y_E}\left[\frac{1}{\pi}\frac{\sin(\Delta)}{E-E'}-\cos(\Delta)\delta(E-E')\right],
  \end{equation}
  \begin{equation}
    c_1^\sigma(E')=\frac{V_{-k',\sigma}}{Y_E}\left[\frac{1}{\pi}\frac{\sin(\Delta)}{E-E'}-\cos(\Delta)\delta(E-E')\right],
  \end{equation}
\end{subequations}
\begin{subequations}
  \begin{equation}
    a_2=0,
  \end{equation}
  \begin{equation}
    b_2^\sigma(E')=\frac{V_{-k',\sigma}}{Y_E}\delta(E-E'),
  \end{equation}
  \begin{equation}
    c_2^\sigma(E')=-\frac{V_{k',\sigma}}{Y_E}\delta(E-E')
  \end{equation}
\end{subequations}
with the following parameters: $\Delta=\arctan[\Gamma/(\tilde E_0-E)]$, $Y_E=\sqrt{V_{k_E,\sigma}^2+V_{-k_E,\sigma}^2}$ (it does not depend on $\sigma$) and $k'=\sqrt{2mE'}$.

We note that
\begin{equation}
  \tilde E_0=E_0+{\rm v.p.}\int\limits_0^\infty\d E'\frac{D(E')Y_{E'}^2}{4(E-E')}
\end{equation}
and
\begin{equation}
  \Gamma=\pi\frac{D(E)Y_E^2}{4}
\end{equation}
in agreement with Eq.~\eqref{eq:Sigma_R}: $\Sigma_\sigma^R(E)=\tilde E_0-E_0-\i\Gamma$.

We also note that these eigenfunctions do not form a complete set, because any potential in one dimensional problem produces a bound state with a negative energy $E_b<0$. The two Kramers degenerate truly localized states have the same form of Eq.~\eqref{eq:Psis}:
\begin{multline}
  \Psi_{0}^\sigma=a_{0}^\sigma\Phi_{\sigma}+{\rm v.p.}\int\limits_0^\infty\d E'\left[b_{0}^\sigma(E')\Psi_{k',-\sigma}\right.\\\left.+c_{0}^\sigma(E')\Psi_{-k',-\sigma}\right].
\end{multline}
Here the coefficients have the form
\begin{equation}
  b_0^\sigma(E')=\frac{D(E')a_0V_{k'}^\sigma}{4(E_b-E')},
  \qquad
  c_0^\sigma(E')=\frac{D(E')a_0V_{-k'}^\sigma}{4(E_b-E')},
\end{equation}
and the coefficient $a_0$ should be determined from the normalization of these wave functions. The energy of these truly bound states can be found from the relation
\begin{equation}
  E_b=E_0+\int_0^\infty\frac{D(E')Y_{E'}^2}{4(E_b-E')}.
\end{equation}
One can see that the Green's function~\eqref{eq:G_R_U0} indeed has a pole at this energy. However, for $E_0\gg\Gamma$ this state is almost delocalized, $a_0^\sigma\ll 1$, so the contribution of this state can be neglected.

For the simple Gaussian wave functions~\eqref{eq:Psi_Gaussian} and matrix elements~\eqref{eq:V_pm} one can readily find the width of the quasi bound state
\begin{multline}
  \label{eq:Gamma_exp}
  \Gamma=\frac{3\sqrt{\pi}\gamma_2^2\hbar^2mV_z^2}{2\sqrt{2}a^2m_0^2}\left[a^3k_E^3+2(3d^2/a^2-1)k_Ea
    \right.\\\left.
    +(d^2/a^2-1)^2/(k_Ea)\right]\exp\left(-k_E^2a^2/2-d^2/a^2\right).
\end{multline}
For the energy renormalization we obtain
\begin{multline}
  \tilde E-E_0=\frac{3\gamma_2^2\hbar^2mV_z^2}{2\sqrt{2}a^2m_0^2}\left\{2\left[(k_Ea)^3+2k_Ea(3d^2/a^2-1)
      \right.\right.\\\left.
        +(d^2/a^2-1)^2/(k_Ea)\right]\mathcal D(k_Ea/\sqrt{2})
      \\\left.
        +\sqrt{2}\left[1-6(d/a)^2-(k_Ea)^2\right]\right\}\exp\left(-d^2/a^2\right),
\end{multline}
where $\mathcal D(x)=e^{-x^{2}}\int_{0}^{x}\e^{y^{2}}\d y$ is the Dawson function.
We remind that $k_E=\sqrt{2mE}/\hbar$ is used for brevity, and here we recovered the reduced Plank constant.

To describe the population of the QD spin states in the presence of the current, we consider the following linear combinations of the eigenfunctions at the given energy:
\begin{subequations}
  \begin{equation}
    \Psi_L^\sigma=\frac{1}{Y_E}\left[V_{-k_E,\sigma}\Psi_2^\sigma-V_{k_E,\sigma}\e^{\i\Delta}\Psi_1^\sigma\right],
  \end{equation}
  \begin{equation}
    \Psi_R^\sigma=\frac{1}{Y_E}\left[V_{k_E,\sigma}\Psi_2^\sigma+V_{-k_E,\sigma}\e^{\i\Delta}\Psi_1^\sigma\right].
  \end{equation}
\end{subequations}
These wave functions have the asymptotic behaviour $\Psi_{L/R}^\sigma\propto\exp(\pm\i k_Ex)$ for $x\to\pm\infty$, respectively. Thus they describe the heavy hole states propagating from the left and right leads, respectively.

The weight of the QD state in these functions is~\cite{PhysRev.124.41}
\begin{equation}
  |a_{L/R,\sigma}^2|=\frac{V_{\pm k_E,\sigma}^2}{(E-\tilde E_0)^2+\Gamma^2}.
\end{equation}
For the given Fermi energies $E_F^L$ and $E_F^R$ in the leads (and low temperatures) the occupancies of the QD states can be found as
\begin{equation}
  \label{eq:n_pm_0_Fano}
  \braket{n_\sigma}=\int\limits_0^{E_F^L}|a_{L,\sigma}^2|\frac{D(E)}{4}\d E+\int\limits_0^{E_F^R}|a_{R,\sigma}^2|\frac{D(E)}{4}\d E,
\end{equation}
which coincides with Eq.~\eqref{eq:n_pm_0} obtained in the Keldysh formalism.

The tunneling matrix elements are exponentially suppressed by the fast decay of the hole wave functions away from the QD and the quantum wire. 
So typically the width of the resonance $\Gamma$ is much smaller than its energy $E_0$. In this case one can use the wide band approximation and neglect the energy dependence of $D(E)$ and $V_{\pm k_E,\sigma}$. Then Eq.~\eqref{eq:n_pm_0_Fano} yields
\begin{multline}
  \label{eq:n_pm_0_wide}
  \braket{n_\pm}=\frac{1}{2}+\frac{1\pm\mathcal C}{2\pi}\arctan\left(\frac{E_F^L-\tilde E_0}{\Gamma}\right)\\
  +\frac{1\mp\mathcal C}{2\pi}\arctan\left(\frac{E_F^R-\tilde E_0}{\Gamma}\right),
\end{multline}
where the chirality $\mathcal C$ is defined in Eq.~\eqref{eq:C_S}, while the quasi bound state energy $\tilde E_0$ and width $\Gamma$ are assumed to be taken at the energy $E_0$.

One can see that the polarization is the largest in the limit of large bias $E_F^L-E_0,E_0-E_F^R\gg\Gamma$. In this limit one has $\braket{n_\pm}=(1\pm\mathcal C)/2$, which yields the polarization
\begin{equation}
  \label{eq:PC}
  P=\mathcal C.
\end{equation}
So the chirality directly defines the largest current induced spin polarization without interaction.

We note that the limit of the large bias can be also described using the phenomenological kinetic equations
\begin{equation}
  \label{eq:kinetic_S0}
  \frac{\d n_\pm}{\d t}=2\Gamma_{L,\pm}-2\Gamma n_\pm,
\end{equation}
which describe the tunneling of the light holes to the QD with the rate $2\Gamma_{L,\pm}$ and out of the QD with the rate $2\Gamma$. In the steady state one obtains once again $n_\pm=\Gamma_{L,\pm}/\Gamma=(1\pm\mathcal C)/2$ and Eq.~\eqref{eq:PC} in agreement with the Keldysh formalism and exact Hamiltonian diagonalization.

\section{S4. Strong Coulomb interaction}

For small quantum dots it is relevant to consider the limit of strong Coulomb interaction, $U\to\infty$. In this limit one can neglect the second term in Eq.~\eqref{eq:Hubbard} for the retarded bare Green's function $G_{0,\sigma}^R(\omega)$ [Eq.~\eqref{eq:G_R} in the main text]:
\begin{equation}
  G_{0,\sigma}^R(\omega)=\frac{1-\braket{n_{-\sigma}}}{\omega-E_0+\i\delta}.
\end{equation}
Similarly to the previous subsection we obtain the dressed retarded Green's function
 \begin{equation}
   G_\sigma^R(\omega)=\frac{1-\braket{n_{-\sigma}}}{\omega-E_{0,\sigma}(\omega)+\i\Gamma_\sigma(\omega)},
 \end{equation}
where $E_{0,\sigma}(\omega)=\braket{n_{-\sigma}}E_0+(1-\braket{n_{-\sigma}})\tilde E_0(\omega)$ and $\Gamma_\sigma(\omega)=(1-\braket{n_{-\sigma}})\Gamma(\omega)$.
Thus we obtain the suppressed amplitude of the Green's function, smaller renormalization of the energy of the quasi bound state, and its smaller width as compared with Eq.~\eqref{eq:G_R_U0}.

Further, the occupancies of the QD spin states can be found from Eq.~\eqref{eq:n_pm} and~\eqref{eq:Gm}. In the wide band approximation in analogy with Eq.~\eqref{eq:n_pm_0_wide} we obtain
\begin{multline}
  \braket{n_\sigma}=(1-\braket{n_{-\sigma}})\left[\frac{1}{2}+\frac{1+\sigma\mathcal C}{2\pi}\arctan\left(\frac{E_F^L-\tilde E_{0,\sigma}}{\Gamma_\sigma}\right)
  \right.\\\left.
  +\frac{1-\sigma\mathcal C}{2\pi}\arctan\left(\frac{E_F^R-\tilde E_{0,\sigma}}{\Gamma_\sigma}\right)\right].
\end{multline}
This represents the set of two equations for $\braket{n_\pm}$, which should be solved self consistently.

In the limit of large bias $E_F^L-E_0,E_0-E_F^R\gg\Gamma$ one obtains
\begin{equation}
  \braket{n_\sigma}=(1-\braket{n_{-\sigma}})(1+\sigma\mathcal C)/2,
\end{equation}
which yields [Eq.~\eqref{eq:P_max} in the main text]:
\begin{equation}
  \label{eq:P_max_S}
  P_{\rm max}=\frac{2\mathcal C}{1+\mathcal C^2}.
\end{equation}
Thus for the chiral quasi bound state, $\mathcal C=\pm1$, in the presence of the interaction the current induced spin polarization reaches 100\%.

This limit can be again described using the phenomenological kinetic equations
\begin{equation}
  \frac{\d n_\sigma}{\d t}=2\Gamma_{L,\sigma}(1-n_{-\sigma})-2\Gamma n_\sigma,
\end{equation}
which is similar to Eq.~\eqref{eq:kinetic_S0}, but accounts for the Coulomb blockade effect. These equations lead to
\begin{equation}
  n_\sigma=\frac{(1\pm\mathcal C)^2}{3+\mathcal C^2},
\end{equation}
which yields again Eq.~\eqref{eq:P_max_S}.

Generally, one can see that the Coulomb interaction increases the polarization degree. Qualitatively, this happens because the presence of a light hole with the given spin in the QD prevents tunneling of the hole with the opposite spin to the QD. As a result the spin polarization degree increases by a factor of $2/(1+\mathcal C^2)$.

It follows from Eq.~\eqref{eq:P_max_S} that it is enough to have quite a moderate chirality $|\mathcal C|>(10-\sqrt{19})/9\approx0.63$ to obtain the spin polarization degree $P_{\rm max}$ larger then $90$\%. The corresponding region of the system parameters is shown in the inset in Fig.~\ref{fig:energy} in the main text.




\begin{thebibliography}{56}%
\makeatletter
\providecommand \@ifxundefined [1]{%
 \@ifx{#1\undefined}
}%
\providecommand \@ifnum [1]{%
 \ifnum #1\expandafter \@firstoftwo
 \else \expandafter \@secondoftwo
 \fi
}%
\providecommand \@ifx [1]{%
 \ifx #1\expandafter \@firstoftwo
 \else \expandafter \@secondoftwo
 \fi
}%
\providecommand \natexlab [1]{#1}%
\providecommand \enquote  [1]{``#1''}%
\providecommand \bibnamefont  [1]{#1}%
\providecommand \bibfnamefont [1]{#1}%
\providecommand \citenamefont [1]{#1}%
\providecommand \href@noop [0]{\@secondoftwo}%
\providecommand \href [0]{\begingroup \@sanitize@url \@href}%
\providecommand \@href[1]{\@@startlink{#1}\@@href}%
\providecommand \@@href[1]{\endgroup#1\@@endlink}%
\providecommand \@sanitize@url [0]{\catcode `\\12\catcode `\$12\catcode
  `\&12\catcode `\#12\catcode `\^12\catcode `\_12\catcode `\%12\relax}%
\providecommand \@@startlink[1]{}%
\providecommand \@@endlink[0]{}%
\providecommand \url  [0]{\begingroup\@sanitize@url \@url }%
\providecommand \@url [1]{\endgroup\@href {#1}{\urlprefix }}%
\providecommand \urlprefix  [0]{URL }%
\providecommand \Eprint [0]{\href }%
\providecommand \doibase [0]{http://dx.doi.org/}%
\providecommand \selectlanguage [0]{\@gobble}%
\providecommand \bibinfo  [0]{\@secondoftwo}%
\providecommand \bibfield  [0]{\@secondoftwo}%
\providecommand \translation [1]{[#1]}%
\providecommand \BibitemOpen [0]{}%
\providecommand \bibitemStop [0]{}%
\providecommand \bibitemNoStop [0]{.\EOS\space}%
\providecommand \EOS [0]{\spacefactor3000\relax}%
\providecommand \BibitemShut  [1]{\csname bibitem#1\endcsname}%
\let\auto@bib@innerbib\@empty
\bibitem [{\citenamefont {Dyakonov}(2020)}]{dyakonov2020will}%
  \BibitemOpen
  \bibfield  {author} {\bibinfo {author} {\bibfnamefont {M.~I.}\ \bibnamefont
  {Dyakonov}},\ }\href@noop {} {\emph {\bibinfo {title} {Will We Ever Have a
  Quantum Computer?}}}\ (\bibinfo  {publisher} {Springer Nature},\ \bibinfo
  {year} {2020})\BibitemShut {NoStop}%
\bibitem [{\citenamefont {Watson}\ \emph {et~al.}(2018)\citenamefont {Watson},
  \citenamefont {Philips}, \citenamefont {Kawakami}, \citenamefont {Ward},
  \citenamefont {Scarlino}, \citenamefont {Veldhorst}, \citenamefont {Savage},
  \citenamefont {Lagally}, \citenamefont {Friesen}, \citenamefont
  {Coppersmith}, \citenamefont {Eriksson},\ and\ \citenamefont
  {Vandersypen}}]{Watson2018}%
  \BibitemOpen
  \bibfield  {author} {\bibinfo {author} {\bibfnamefont {T.~F.}\ \bibnamefont
  {Watson}}, \bibinfo {author} {\bibfnamefont {S.~G.~J.}\ \bibnamefont
  {Philips}}, \bibinfo {author} {\bibfnamefont {E.}~\bibnamefont {Kawakami}},
  \bibinfo {author} {\bibfnamefont {D.~R.}\ \bibnamefont {Ward}}, \bibinfo
  {author} {\bibfnamefont {P.}~\bibnamefont {Scarlino}}, \bibinfo {author}
  {\bibfnamefont {M.}~\bibnamefont {Veldhorst}}, \bibinfo {author}
  {\bibfnamefont {D.~E.}\ \bibnamefont {Savage}}, \bibinfo {author}
  {\bibfnamefont {M.~G.}\ \bibnamefont {Lagally}}, \bibinfo {author}
  {\bibfnamefont {M.}~\bibnamefont {Friesen}}, \bibinfo {author} {\bibfnamefont
  {S.~N.}\ \bibnamefont {Coppersmith}}, \bibinfo {author} {\bibfnamefont
  {M.~A.}\ \bibnamefont {Eriksson}}, \ and\ \bibinfo {author} {\bibfnamefont
  {L.~M.~K.}\ \bibnamefont {Vandersypen}},\ }\bibfield  {title} {\enquote
  {\bibinfo {title} {A programmable two-qubit quantum processor in silicon},}\
  }\href {https://doi.org/10.1038/nature25766} {\bibfield  {journal} {\bibinfo
  {journal} {Nature}\ }\textbf {\bibinfo {volume} {555}},\ \bibinfo {pages}
  {633} (\bibinfo {year} {2018})}\BibitemShut {NoStop}%
\bibitem [{\citenamefont {Yoneda}\ \emph {et~al.}(2018)\citenamefont {Yoneda},
  \citenamefont {Takeda}, \citenamefont {Otsuka}, \citenamefont {Nakajima},
  \citenamefont {Delbecq}, \citenamefont {Allison}, \citenamefont {Honda},
  \citenamefont {Kodera}, \citenamefont {Oda}, \citenamefont {Hoshi},
  \citenamefont {Usami}, \citenamefont {Itoh},\ and\ \citenamefont
  {Tarucha}}]{Yoneda2018}%
  \BibitemOpen
  \bibfield  {author} {\bibinfo {author} {\bibfnamefont {J.}~\bibnamefont
  {Yoneda}}, \bibinfo {author} {\bibfnamefont {K.}~\bibnamefont {Takeda}},
  \bibinfo {author} {\bibfnamefont {T.}~\bibnamefont {Otsuka}}, \bibinfo
  {author} {\bibfnamefont {T.}~\bibnamefont {Nakajima}}, \bibinfo {author}
  {\bibfnamefont {M.~R.}\ \bibnamefont {Delbecq}}, \bibinfo {author}
  {\bibfnamefont {G.}~\bibnamefont {Allison}}, \bibinfo {author} {\bibfnamefont
  {T.}~\bibnamefont {Honda}}, \bibinfo {author} {\bibfnamefont
  {T.}~\bibnamefont {Kodera}}, \bibinfo {author} {\bibfnamefont
  {S.}~\bibnamefont {Oda}}, \bibinfo {author} {\bibfnamefont {Y.}~\bibnamefont
  {Hoshi}}, \bibinfo {author} {\bibfnamefont {N.}~\bibnamefont {Usami}},
  \bibinfo {author} {\bibfnamefont {K.~M.}\ \bibnamefont {Itoh}}, \ and\
  \bibinfo {author} {\bibfnamefont {S.}~\bibnamefont {Tarucha}},\ }\bibfield
  {title} {\enquote {\bibinfo {title} {A quantum-dot spin qubit with coherence
  limited by charge noise and fidelity higher than 99.9\%},}\ }\href
  {https://doi.org/10.1038/s41565-017-0014-x} {\bibfield  {journal} {\bibinfo
  {journal} {Nat. Nanotechnol.}\ }\textbf {\bibinfo {volume} {13}},\ \bibinfo
  {pages} {102} (\bibinfo {year} {2018})}\BibitemShut {NoStop}%
\bibitem [{\citenamefont {Yang}\ \emph {et~al.}(2020)\citenamefont {Yang},
  \citenamefont {Leon}, \citenamefont {Hwang}, \citenamefont {Saraiva},
  \citenamefont {Tanttu}, \citenamefont {Huang}, \citenamefont {{Camirand
  Lemyre}}, \citenamefont {Chan}, \citenamefont {Tan}, \citenamefont {Hudson},
  \citenamefont {Itoh}, \citenamefont {Morello}, \citenamefont
  {Pioro-Ladri\'ere}, \citenamefont {Laucht},\ and\ \citenamefont
  {Dzurak}}]{Yang2020}%
  \BibitemOpen
  \bibfield  {author} {\bibinfo {author} {\bibfnamefont {C.~H.}\ \bibnamefont
  {Yang}}, \bibinfo {author} {\bibfnamefont {R.~C.~C.}\ \bibnamefont {Leon}},
  \bibinfo {author} {\bibfnamefont {J.~C.~C.}\ \bibnamefont {Hwang}}, \bibinfo
  {author} {\bibfnamefont {A.}~\bibnamefont {Saraiva}}, \bibinfo {author}
  {\bibfnamefont {T.}~\bibnamefont {Tanttu}}, \bibinfo {author} {\bibfnamefont
  {W.}~\bibnamefont {Huang}}, \bibinfo {author} {\bibfnamefont
  {J.}~\bibnamefont {{Camirand Lemyre}}}, \bibinfo {author} {\bibfnamefont
  {K.~W.}\ \bibnamefont {Chan}}, \bibinfo {author} {\bibfnamefont {K.~Y.}\
  \bibnamefont {Tan}}, \bibinfo {author} {\bibfnamefont {F.~E.}\ \bibnamefont
  {Hudson}}, \bibinfo {author} {\bibfnamefont {K.~M.}\ \bibnamefont {Itoh}},
  \bibinfo {author} {\bibfnamefont {A.}~\bibnamefont {Morello}}, \bibinfo
  {author} {\bibfnamefont {M.}~\bibnamefont {Pioro-Ladri\'ere}}, \bibinfo
  {author} {\bibfnamefont {A.}~\bibnamefont {Laucht}}, \ and\ \bibinfo {author}
  {\bibfnamefont {A.~S.}\ \bibnamefont {Dzurak}},\ }\bibfield  {title}
  {\enquote {\bibinfo {title} {Operation of a silicon quantum processor unit
  cell above one kelvin},}\ }\href {https://doi.org/10.1038/s41586-020-2171-6}
  {\bibfield  {journal} {\bibinfo  {journal} {Nature}\ }\textbf {\bibinfo
  {volume} {580}},\ \bibinfo {pages} {350} (\bibinfo {year}
  {2020})}\BibitemShut {NoStop}%
\bibitem [{\citenamefont {Nowack}\ \emph {et~al.}(2007)\citenamefont {Nowack},
  \citenamefont {Koppens}, \citenamefont {Nazarov},\ and\ \citenamefont
  {Vandersypen}}]{Nowack07}%
  \BibitemOpen
  \bibfield  {author} {\bibinfo {author} {\bibfnamefont {K.~C.}\ \bibnamefont
  {Nowack}}, \bibinfo {author} {\bibfnamefont {F.~H.~L.}\ \bibnamefont
  {Koppens}}, \bibinfo {author} {\bibfnamefont {Y.~V.}\ \bibnamefont
  {Nazarov}}, \ and\ \bibinfo {author} {\bibfnamefont {L.~M.~K.}\ \bibnamefont
  {Vandersypen}},\ }\bibfield  {title} {\enquote {\bibinfo {title} {Coherent
  Control of a Single Electron Spin with Electric Fields},}\ }\href {\doibase
  10.1126/science.1148092} {\bibfield  {journal} {\bibinfo  {journal}
  {Science}\ }\textbf {\bibinfo {volume} {318}},\ \bibinfo {pages} {1430}
  (\bibinfo {year} {2007})}\BibitemShut {NoStop}%
\bibitem [{\citenamefont {Zajac}\ \emph {et~al.}(2018)\citenamefont {Zajac},
  \citenamefont {Sigillito}, \citenamefont {Russ}, \citenamefont {Borjans},
  \citenamefont {Taylor}, \citenamefont {Burkard},\ and\ \citenamefont
  {Petta}}]{Zajac439}%
  \BibitemOpen
  \bibfield  {author} {\bibinfo {author} {\bibfnamefont {D.~M.}\ \bibnamefont
  {Zajac}}, \bibinfo {author} {\bibfnamefont {A.~J.}\ \bibnamefont
  {Sigillito}}, \bibinfo {author} {\bibfnamefont {M.}~\bibnamefont {Russ}},
  \bibinfo {author} {\bibfnamefont {F.}~\bibnamefont {Borjans}}, \bibinfo
  {author} {\bibfnamefont {J.~M.}\ \bibnamefont {Taylor}}, \bibinfo {author}
  {\bibfnamefont {G.}~\bibnamefont {Burkard}}, \ and\ \bibinfo {author}
  {\bibfnamefont {J.~R.}\ \bibnamefont {Petta}},\ }\bibfield  {title} {\enquote
  {\bibinfo {title} {Resonantly driven CNOT gate for electron spins},}\ }\href
  {\doibase 10.1126/science.aao5965} {\bibfield  {journal} {\bibinfo  {journal}
  {Science}\ }\textbf {\bibinfo {volume} {359}},\ \bibinfo {pages} {439}
  (\bibinfo {year} {2018})}\BibitemShut {NoStop}%
\bibitem [{\citenamefont {Basso~Basset}\ \emph {et~al.}(2019)\citenamefont
  {Basso~Basset}, \citenamefont {Rota}, \citenamefont {Schimpf}, \citenamefont
  {Tedeschi}, \citenamefont {Zeuner}, \citenamefont {Covre~da Silva},
  \citenamefont {Reindl}, \citenamefont {Zwiller}, \citenamefont {J\"ons},
  \citenamefont {Rastelli},\ and\ \citenamefont
  {Trotta}}]{PhysRevLett.123.160501}%
  \BibitemOpen
  \bibfield  {author} {\bibinfo {author} {\bibfnamefont {F.}~\bibnamefont
  {Basso~Basset}}, \bibinfo {author} {\bibfnamefont {M.~B.}\ \bibnamefont
  {Rota}}, \bibinfo {author} {\bibfnamefont {C.}~\bibnamefont {Schimpf}},
  \bibinfo {author} {\bibfnamefont {D.}~\bibnamefont {Tedeschi}}, \bibinfo
  {author} {\bibfnamefont {K.~D.}\ \bibnamefont {Zeuner}}, \bibinfo {author}
  {\bibfnamefont {S.~F.}\ \bibnamefont {Covre~da Silva}}, \bibinfo {author}
  {\bibfnamefont {M.}~\bibnamefont {Reindl}}, \bibinfo {author} {\bibfnamefont
  {V.}~\bibnamefont {Zwiller}}, \bibinfo {author} {\bibfnamefont {K.~D.}\
  \bibnamefont {J\"ons}}, \bibinfo {author} {\bibfnamefont {A.}~\bibnamefont
  {Rastelli}}, \ and\ \bibinfo {author} {\bibfnamefont {R.}~\bibnamefont
  {Trotta}},\ }\bibfield  {title} {\enquote {\bibinfo {title} {Entanglement
  Swapping with Photons Generated on Demand by a Quantum Dot},}\ }\href
  {\doibase 10.1103/PhysRevLett.123.160501} {\bibfield  {journal} {\bibinfo
  {journal} {Phys. Rev. Lett.}\ }\textbf {\bibinfo {volume} {123}},\ \bibinfo
  {pages} {160501} (\bibinfo {year} {2019})}\BibitemShut {NoStop}%
\bibitem [{\citenamefont {Mills}\ \emph {et~al.}(2019)\citenamefont {Mills},
  \citenamefont {Zajac}, \citenamefont {Gullans}, \citenamefont {Schupp},
  \citenamefont {Hazard},\ and\ \citenamefont {Petta}}]{Mills2019}%
  \BibitemOpen
  \bibfield  {author} {\bibinfo {author} {\bibfnamefont {A.~R.}\ \bibnamefont
  {Mills}}, \bibinfo {author} {\bibfnamefont {D.~M.}\ \bibnamefont {Zajac}},
  \bibinfo {author} {\bibfnamefont {M.~J.}\ \bibnamefont {Gullans}}, \bibinfo
  {author} {\bibfnamefont {F.~J.}\ \bibnamefont {Schupp}}, \bibinfo {author}
  {\bibfnamefont {T.~M.}\ \bibnamefont {Hazard}}, \ and\ \bibinfo {author}
  {\bibfnamefont {J.~R.}\ \bibnamefont {Petta}},\ }\bibfield  {title} {\enquote
  {\bibinfo {title} {Shuttling a single charge across a one-dimensional array
  of silicon quantum dots},}\ }\href
  {https://doi.org/10.1038/s41467-019-08970-z} {\bibfield  {journal} {\bibinfo
  {journal} {Nat. Commun.}\ }\textbf {\bibinfo {volume} {10}},\ \bibinfo
  {pages} {1063} (\bibinfo {year} {2019})}\BibitemShut {NoStop}%
\bibitem [{\citenamefont {Qiao}\ \emph {et~al.}(2021)\citenamefont {Qiao},
  \citenamefont {Kandel}, \citenamefont {Fallahi}, \citenamefont {Gardner},
  \citenamefont {Manfra}, \citenamefont {Hu},\ and\ \citenamefont
  {Nichol}}]{PhysRevLett.126.017701}%
  \BibitemOpen
  \bibfield  {author} {\bibinfo {author} {\bibfnamefont {H.}~\bibnamefont
  {Qiao}}, \bibinfo {author} {\bibfnamefont {Y.~P.}\ \bibnamefont {Kandel}},
  \bibinfo {author} {\bibfnamefont {S.}~\bibnamefont {Fallahi}}, \bibinfo
  {author} {\bibfnamefont {G.~C.}\ \bibnamefont {Gardner}}, \bibinfo {author}
  {\bibfnamefont {M.~J.}\ \bibnamefont {Manfra}}, \bibinfo {author}
  {\bibfnamefont {X.}~\bibnamefont {Hu}}, \ and\ \bibinfo {author}
  {\bibfnamefont {J.~M.}\ \bibnamefont {Nichol}},\ }\bibfield  {title}
  {\enquote {\bibinfo {title} {Long-Distance Superexchange between
  Semiconductor Quantum-Dot Electron Spins},}\ }\href {\doibase
  10.1103/PhysRevLett.126.017701} {\bibfield  {journal} {\bibinfo  {journal}
  {Phys. Rev. Lett.}\ }\textbf {\bibinfo {volume} {126}},\ \bibinfo {pages}
  {017701} (\bibinfo {year} {2021})}\BibitemShut {NoStop}%
\bibitem [{\citenamefont {Carter}\ \emph {et~al.}(2021)\citenamefont {Carter},
  \citenamefont {Badescu}, \citenamefont {Bracker}, \citenamefont {Yakes},
  \citenamefont {Tran}, \citenamefont {Grim},\ and\ \citenamefont
  {Gammon}}]{PhysRevLett.126.107401}%
  \BibitemOpen
  \bibfield  {author} {\bibinfo {author} {\bibfnamefont {S.~G.}\ \bibnamefont
  {Carter}}, \bibinfo {author} {\bibfnamefont {S.~C.}\ \bibnamefont {Badescu}},
  \bibinfo {author} {\bibfnamefont {A.~S.}\ \bibnamefont {Bracker}}, \bibinfo
  {author} {\bibfnamefont {M.~K.}\ \bibnamefont {Yakes}}, \bibinfo {author}
  {\bibfnamefont {K.~X.}\ \bibnamefont {Tran}}, \bibinfo {author}
  {\bibfnamefont {J.~Q.}\ \bibnamefont {Grim}}, \ and\ \bibinfo {author}
  {\bibfnamefont {D.}~\bibnamefont {Gammon}},\ }\bibfield  {title} {\enquote
  {\bibinfo {title} {Coherent Population Trapping Combined with Cycling
  Transitions for Quantum Dot Hole Spins Using Triplet Trion States},}\ }\href
  {\doibase 10.1103/PhysRevLett.126.107401} {\bibfield  {journal} {\bibinfo
  {journal} {Phys. Rev. Lett.}\ }\textbf {\bibinfo {volume} {126}},\ \bibinfo
  {pages} {107401} (\bibinfo {year} {2021})}\BibitemShut {NoStop}%
\bibitem [{\citenamefont {Ivchenko}\ and\ \citenamefont
  {Pikus}(1978)}]{ivchenko1978new}%
  \BibitemOpen
  \bibfield  {author} {\bibinfo {author} {\bibfnamefont {E.~L.}\ \bibnamefont
  {Ivchenko}}\ and\ \bibinfo {author} {\bibfnamefont {G.~E.}\ \bibnamefont
  {Pikus}},\ }\bibfield  {title} {\enquote {\bibinfo {title} {New photogalvanic
  effect in gyrotropic crystals},}\ }\href@noop {} {\bibfield  {journal}
  {\bibinfo  {journal} {JETP Lett.}\ }\textbf {\bibinfo {volume} {27}},\
  \bibinfo {pages} {604} (\bibinfo {year} {1978})}\BibitemShut {NoStop}%
\bibitem [{\citenamefont {Vorob'ev}\ \emph {et~al.}(1979)\citenamefont
  {Vorob'ev}, \citenamefont {Ivchenko}, \citenamefont {Pikus}, \citenamefont
  {Farbshtein}, \citenamefont {Shalygin},\ and\ \citenamefont
  {Shturbin}}]{vorob1979optical}%
  \BibitemOpen
  \bibfield  {author} {\bibinfo {author} {\bibfnamefont {L.~E.}\ \bibnamefont
  {Vorob'ev}}, \bibinfo {author} {\bibfnamefont {E.~L.}\ \bibnamefont
  {Ivchenko}}, \bibinfo {author} {\bibfnamefont {G.~E.}\ \bibnamefont {Pikus}},
  \bibinfo {author} {\bibfnamefont {I.~I.}\ \bibnamefont {Farbshtein}},
  \bibinfo {author} {\bibfnamefont {V.~A.}\ \bibnamefont {Shalygin}}, \ and\
  \bibinfo {author} {\bibfnamefont {A.~V.}\ \bibnamefont {Shturbin}},\
  }\bibfield  {title} {\enquote {\bibinfo {title} {Optical activity in
  tellurium induced by a current},}\ }\href@noop {} {\bibfield  {journal}
  {\bibinfo  {journal} {JETP Lett.}\ }\textbf {\bibinfo {volume} {29}},\
  \bibinfo {pages} {441} (\bibinfo {year} {1979})}\BibitemShut {NoStop}%
\bibitem [{\citenamefont {Ganichev}\ \emph {et~al.}(2006)\citenamefont
  {Ganichev}, \citenamefont {Danilov}, \citenamefont {Schneider}, \citenamefont
  {Bel'kov}, \citenamefont {Golub}, \citenamefont {Wegscheider}, \citenamefont
  {Weiss},\ and\ \citenamefont {Prettl}}]{Ganichev_110}%
  \BibitemOpen
  \bibfield  {author} {\bibinfo {author} {\bibfnamefont {S.~D.}\ \bibnamefont
  {Ganichev}}, \bibinfo {author} {\bibfnamefont {S.~N.}\ \bibnamefont
  {Danilov}}, \bibinfo {author} {\bibfnamefont {P.}~\bibnamefont {Schneider}},
  \bibinfo {author} {\bibfnamefont {V.~V.}\ \bibnamefont {Bel'kov}}, \bibinfo
  {author} {\bibfnamefont {L.~E.}\ \bibnamefont {Golub}}, \bibinfo {author}
  {\bibfnamefont {W.}~\bibnamefont {Wegscheider}}, \bibinfo {author}
  {\bibfnamefont {D.}~\bibnamefont {Weiss}}, \ and\ \bibinfo {author}
  {\bibfnamefont {W.}~\bibnamefont {Prettl}},\ }\bibfield  {title} {\enquote
  {\bibinfo {title} {Electric current-induced spin orientation in quantum well
  structures},}\ }\href {\doibase 10.1016/j.jmmm.2005.10.048} {\bibfield
  {journal} {\bibinfo  {journal} {J. Magn. Magn. Mater.}\ }\textbf {\bibinfo
  {volume} {300}},\ \bibinfo {pages} {127} (\bibinfo {year}
  {2006})}\BibitemShut {NoStop}%
\bibitem [{\citenamefont {Silov}\ \emph {et~al.}(2004)\citenamefont {Silov},
  \citenamefont {Blajnov}, \citenamefont {Wolter}, \citenamefont {Hey},
  \citenamefont {Ploog},\ and\ \citenamefont {Averkiev}}]{silov04}%
  \BibitemOpen
  \bibfield  {author} {\bibinfo {author} {\bibfnamefont {A.~Y.}\ \bibnamefont
  {Silov}}, \bibinfo {author} {\bibfnamefont {P.~A.}\ \bibnamefont {Blajnov}},
  \bibinfo {author} {\bibfnamefont {J.~H.}\ \bibnamefont {Wolter}}, \bibinfo
  {author} {\bibfnamefont {R.}~\bibnamefont {Hey}}, \bibinfo {author}
  {\bibfnamefont {K.~H.}\ \bibnamefont {Ploog}}, \ and\ \bibinfo {author}
  {\bibfnamefont {N.~S.}\ \bibnamefont {Averkiev}},\ }\bibfield  {title}
  {\enquote {\bibinfo {title} {Current-induced spin polarization at a single
  heterojunction},}\ }\href {\doibase 10.1063/1.1833565} {\bibfield  {journal}
  {\bibinfo  {journal} {Appl. Phys. Lett.}\ }\textbf {\bibinfo {volume} {85}},\
  \bibinfo {pages} {5929} (\bibinfo {year} {2004})}\BibitemShut {NoStop}%
\bibitem [{\citenamefont {Kato}\ \emph {et~al.}(2004)\citenamefont {Kato},
  \citenamefont {Myers}, \citenamefont {Gossard},\ and\ \citenamefont
  {Awschalom}}]{PhysRevLett.93.176601}%
  \BibitemOpen
  \bibfield  {author} {\bibinfo {author} {\bibfnamefont {Y.~K.}\ \bibnamefont
  {Kato}}, \bibinfo {author} {\bibfnamefont {R.~C.}\ \bibnamefont {Myers}},
  \bibinfo {author} {\bibfnamefont {A.~C.}\ \bibnamefont {Gossard}}, \ and\
  \bibinfo {author} {\bibfnamefont {D.~D.}\ \bibnamefont {Awschalom}},\
  }\bibfield  {title} {\enquote {\bibinfo {title} {Current-Induced Spin
  Polarization in Strained Semiconductors},}\ }\href {\doibase
  10.1103/PhysRevLett.93.176601} {\bibfield  {journal} {\bibinfo  {journal}
  {Phys. Rev. Lett.}\ }\textbf {\bibinfo {volume} {93}},\ \bibinfo {pages}
  {176601} (\bibinfo {year} {2004})}\BibitemShut {NoStop}%
\bibitem [{\citenamefont {Norman}\ \emph {et~al.}(2014)\citenamefont {Norman},
  \citenamefont {Trowbridge}, \citenamefont {Awschalom},\ and\ \citenamefont
  {Sih}}]{Norman2014}%
  \BibitemOpen
  \bibfield  {author} {\bibinfo {author} {\bibfnamefont {B.~M.}\ \bibnamefont
  {Norman}}, \bibinfo {author} {\bibfnamefont {C.~J.}\ \bibnamefont
  {Trowbridge}}, \bibinfo {author} {\bibfnamefont {D.~D.}\ \bibnamefont
  {Awschalom}}, \ and\ \bibinfo {author} {\bibfnamefont {V.}~\bibnamefont
  {Sih}},\ }\bibfield  {title} {\enquote {\bibinfo {title} {Current-Induced
  Spin Polarization in Anisotropic Spin-Orbit Fields},}\ }\href {\doibase
  10.1103/PhysRevLett.112.056601} {\bibfield  {journal} {\bibinfo  {journal}
  {Phys. Rev. Lett.}\ }\textbf {\bibinfo {volume} {112}},\ \bibinfo {pages}
  {056601} (\bibinfo {year} {2014})}\BibitemShut {NoStop}%
\bibitem [{\citenamefont {Stepanov}\ \emph {et~al.}(2014)\citenamefont
  {Stepanov}, \citenamefont {Kuhlen}, \citenamefont {Ersfeld}, \citenamefont
  {Lepsa},\ and\ \citenamefont {Beschoten}}]{doi:10.1063/1.4864468}%
  \BibitemOpen
  \bibfield  {author} {\bibinfo {author} {\bibfnamefont {I.}~\bibnamefont
  {Stepanov}}, \bibinfo {author} {\bibfnamefont {S.}~\bibnamefont {Kuhlen}},
  \bibinfo {author} {\bibfnamefont {M.}~\bibnamefont {Ersfeld}}, \bibinfo
  {author} {\bibfnamefont {M.}~\bibnamefont {Lepsa}}, \ and\ \bibinfo {author}
  {\bibfnamefont {B.}~\bibnamefont {Beschoten}},\ }\bibfield  {title} {\enquote
  {\bibinfo {title} {All-electrical time-resolved spin generation and spin
  manipulation in n-InGaAs},}\ }\href {\doibase 10.1063/1.4864468} {\bibfield
  {journal} {\bibinfo  {journal} {Appl. Phys. Lett.}\ }\textbf {\bibinfo
  {volume} {104}},\ \bibinfo {pages} {062406} (\bibinfo {year}
  {2014})}\BibitemShut {NoStop}%
\bibitem [{\citenamefont {Ganichev}\ \emph {et~al.}(2012)\citenamefont
  {Ganichev}, \citenamefont {Trushin},\ and\ \citenamefont
  {Schliemann}}]{ganichev2012spin}%
  \BibitemOpen
  \bibfield  {author} {\bibinfo {author} {\bibfnamefont {S.~D.}\ \bibnamefont
  {Ganichev}}, \bibinfo {author} {\bibfnamefont {M.}~\bibnamefont {Trushin}}, \
  and\ \bibinfo {author} {\bibfnamefont {J.}~\bibnamefont {Schliemann}},\
  }\href@noop {} {\enquote {\bibinfo {title} {Spin polarization by current in
  Handbook of Spin Transport and Magnetism, edited by E. Y, Tsymbal and I.
  Zutic, p. 487},}\ } (\bibinfo {year} {Chapman \& Hall, Boca Raton,
  2012})\BibitemShut {NoStop}%
\bibitem [{\citenamefont {Golub}\ and\ \citenamefont
  {Ivchenko}(2013)}]{Golub2013}%
  \BibitemOpen
  \bibfield  {author} {\bibinfo {author} {\bibfnamefont {L.~E.}\ \bibnamefont
  {Golub}}\ and\ \bibinfo {author} {\bibfnamefont {E.~L.}\ \bibnamefont
  {Ivchenko}},\ }\bibfield  {title} {\enquote {\bibinfo {title} {Spin-dependent
  phenomena in semiconductors in strong electric fields},}\ }\href {\doibase
  10.1088/1367-2630/15/12/125003} {\bibfield  {journal} {\bibinfo  {journal}
  {New J. Phys.}\ }\textbf {\bibinfo {volume} {15}},\ \bibinfo {pages} {125003}
  (\bibinfo {year} {2013})}\BibitemShut {NoStop}%
\bibitem [{\citenamefont {Smirnov}\ and\ \citenamefont
  {Golub}(2017)}]{Hopping_spin}%
  \BibitemOpen
  \bibfield  {author} {\bibinfo {author} {\bibfnamefont {D.~S.}\ \bibnamefont
  {Smirnov}}\ and\ \bibinfo {author} {\bibfnamefont {L.~E.}\ \bibnamefont
  {Golub}},\ }\bibfield  {title} {\enquote {\bibinfo {title} {Electrical Spin
  Orientation, Spin-Galvanic, and Spin-Hall Effects in Disordered
  Two-Dimensional Systems},}\ }\href {\doibase 10.1103/PhysRevLett.118.116801}
  {\bibfield  {journal} {\bibinfo  {journal} {Phys. Rev. Lett.}\ }\textbf
  {\bibinfo {volume} {118}},\ \bibinfo {pages} {116801} (\bibinfo {year}
  {2017})}\BibitemShut {NoStop}%
\bibitem [{\citenamefont {Huber}\ \emph {et~al.}(2017)\citenamefont {Huber},
  \citenamefont {Reindl}, \citenamefont {Huo}, \citenamefont {Huang},
  \citenamefont {Wildmann}, \citenamefont {Schmidt}, \citenamefont {Rastelli},\
  and\ \citenamefont {Trotta}}]{Huber2017}%
  \BibitemOpen
  \bibfield  {author} {\bibinfo {author} {\bibfnamefont {D.}~\bibnamefont
  {Huber}}, \bibinfo {author} {\bibfnamefont {M.}~\bibnamefont {Reindl}},
  \bibinfo {author} {\bibfnamefont {Y.}~\bibnamefont {Huo}}, \bibinfo {author}
  {\bibfnamefont {H.}~\bibnamefont {Huang}}, \bibinfo {author} {\bibfnamefont
  {J.~S.}\ \bibnamefont {Wildmann}}, \bibinfo {author} {\bibfnamefont {O.~G.}\
  \bibnamefont {Schmidt}}, \bibinfo {author} {\bibfnamefont {A.}~\bibnamefont
  {Rastelli}}, \ and\ \bibinfo {author} {\bibfnamefont {R.}~\bibnamefont
  {Trotta}},\ }\bibfield  {title} {\enquote {\bibinfo {title} {Highly
  indistinguishable and strongly entangled photons from symmetric GaAs quantum
  dots},}\ }\href {https://doi.org/10.1038/ncomms15506} {\bibfield  {journal}
  {\bibinfo  {journal} {Nat. Commun.}\ }\textbf {\bibinfo {volume} {8}},\
  \bibinfo {pages} {15506} (\bibinfo {year} {2017})}\BibitemShut {NoStop}%
\bibitem [{\citenamefont {Chekhovich}\ \emph {et~al.}(2020)\citenamefont
  {Chekhovich}, \citenamefont {da~Silva},\ and\ \citenamefont
  {Rastelli}}]{Chekhovich_register}%
  \BibitemOpen
  \bibfield  {author} {\bibinfo {author} {\bibfnamefont {E.~A.}\ \bibnamefont
  {Chekhovich}}, \bibinfo {author} {\bibfnamefont {S.~F.~C.}\ \bibnamefont
  {da~Silva}}, \ and\ \bibinfo {author} {\bibfnamefont {A.}~\bibnamefont
  {Rastelli}},\ }\bibfield  {title} {\enquote {\bibinfo {title} {Nuclear spin
  quantum register in an optically active semiconductor quantum dot},}\ }\href
  {https://doi.org/10.1038/s41565-020-0769-3} {\bibfield  {journal} {\bibinfo
  {journal} {Nat. nanotech.}\ }\textbf {\bibinfo {volume} {15}},\ \bibinfo
  {pages} {999} (\bibinfo {year} {2020})}\BibitemShut {NoStop}%
\bibitem [{\citenamefont {Najer}\ \emph {et~al.}(2019)\citenamefont {Najer},
  \citenamefont {S{\"o}llner}, \citenamefont {Sekatski}, \citenamefont
  {Dolique}, \citenamefont {L{\"o}bl}, \citenamefont {Riedel}, \citenamefont
  {Schott}, \citenamefont {Starosielec}, \citenamefont {Valentin},
  \citenamefont {Wieck}, \citenamefont {Sangouard}, \citenamefont {Ludwig},\
  and\ \citenamefont {Warburton}}]{Strong_coup_exp}%
  \BibitemOpen
  \bibfield  {author} {\bibinfo {author} {\bibfnamefont {D.}~\bibnamefont
  {Najer}}, \bibinfo {author} {\bibfnamefont {I.}~\bibnamefont {S{\"o}llner}},
  \bibinfo {author} {\bibfnamefont {P.}~\bibnamefont {Sekatski}}, \bibinfo
  {author} {\bibfnamefont {V.}~\bibnamefont {Dolique}}, \bibinfo {author}
  {\bibfnamefont {M.~C.}\ \bibnamefont {L{\"o}bl}}, \bibinfo {author}
  {\bibfnamefont {D.}~\bibnamefont {Riedel}}, \bibinfo {author} {\bibfnamefont
  {R.}~\bibnamefont {Schott}}, \bibinfo {author} {\bibfnamefont
  {S.}~\bibnamefont {Starosielec}}, \bibinfo {author} {\bibfnamefont {S.~R.}\
  \bibnamefont {Valentin}}, \bibinfo {author} {\bibfnamefont {A.~D.}\
  \bibnamefont {Wieck}}, \bibinfo {author} {\bibfnamefont {N.}~\bibnamefont
  {Sangouard}}, \bibinfo {author} {\bibfnamefont {A.}~\bibnamefont {Ludwig}}, \
  and\ \bibinfo {author} {\bibfnamefont {R.~J.}\ \bibnamefont {Warburton}},\
  }\bibfield  {title} {\enquote {\bibinfo {title} {A gated quantum dot strongly
  coupled to an optical microcavity},}\ }\href {\doibase
  10.1038/s41586-019-1709-y} {\bibfield  {journal} {\bibinfo  {journal}
  {Nature}\ }\textbf {\bibinfo {volume} {575}},\ \bibinfo {pages} {622}
  (\bibinfo {year} {2019})}\BibitemShut {NoStop}%
\bibitem [{\citenamefont {Veldhorst}\ \emph {et~al.}(2015)\citenamefont
  {Veldhorst}, \citenamefont {Yang}, \citenamefont {Hwang}, \citenamefont
  {Huang}, \citenamefont {Dehollain}, \citenamefont {Muhonen}, \citenamefont
  {Simmons}, \citenamefont {Laucht}, \citenamefont {Hudson}, \citenamefont
  {Itoh}, \citenamefont {Morello},\ and\ \citenamefont
  {Dzurak}}]{Veldhorst2015}%
  \BibitemOpen
  \bibfield  {author} {\bibinfo {author} {\bibfnamefont {M.}~\bibnamefont
  {Veldhorst}}, \bibinfo {author} {\bibfnamefont {C.~H.}\ \bibnamefont {Yang}},
  \bibinfo {author} {\bibfnamefont {J.~C.~C.}\ \bibnamefont {Hwang}}, \bibinfo
  {author} {\bibfnamefont {W.}~\bibnamefont {Huang}}, \bibinfo {author}
  {\bibfnamefont {J.~P.}\ \bibnamefont {Dehollain}}, \bibinfo {author}
  {\bibfnamefont {J.~T.}\ \bibnamefont {Muhonen}}, \bibinfo {author}
  {\bibfnamefont {S.}~\bibnamefont {Simmons}}, \bibinfo {author} {\bibfnamefont
  {A.}~\bibnamefont {Laucht}}, \bibinfo {author} {\bibfnamefont {F.~E.}\
  \bibnamefont {Hudson}}, \bibinfo {author} {\bibfnamefont {K.~M.}\
  \bibnamefont {Itoh}}, \bibinfo {author} {\bibfnamefont {A.}~\bibnamefont
  {Morello}}, \ and\ \bibinfo {author} {\bibfnamefont {A.~S.}\ \bibnamefont
  {Dzurak}},\ }\bibfield  {title} {\enquote {\bibinfo {title} {A two-qubit
  logic gate in silicon},}\ }\href {https://doi.org/10.1038/nature15263}
  {\bibfield  {journal} {\bibinfo  {journal} {Nature}\ }\textbf {\bibinfo
  {volume} {526}},\ \bibinfo {pages} {410} (\bibinfo {year}
  {2015})}\BibitemShut {NoStop}%
\bibitem [{\citenamefont {Mi}\ \emph {et~al.}(2017)\citenamefont {Mi},
  \citenamefont {Cady}, \citenamefont {Zajac}, \citenamefont {Deelman},\ and\
  \citenamefont {Petta}}]{Mi156}%
  \BibitemOpen
  \bibfield  {author} {\bibinfo {author} {\bibfnamefont {X.}~\bibnamefont
  {Mi}}, \bibinfo {author} {\bibfnamefont {J.~V.}\ \bibnamefont {Cady}},
  \bibinfo {author} {\bibfnamefont {D.~M.}\ \bibnamefont {Zajac}}, \bibinfo
  {author} {\bibfnamefont {P.~W.}\ \bibnamefont {Deelman}}, \ and\ \bibinfo
  {author} {\bibfnamefont {J.~R.}\ \bibnamefont {Petta}},\ }\bibfield  {title}
  {\enquote {\bibinfo {title} {Strong coupling of a single electron in silicon
  to a microwave photon},}\ }\href {\doibase 10.1126/science.aal2469}
  {\bibfield  {journal} {\bibinfo  {journal} {Science}\ }\textbf {\bibinfo
  {volume} {355}},\ \bibinfo {pages} {156} (\bibinfo {year}
  {2017})}\BibitemShut {NoStop}%
\bibitem [{\citenamefont {Borjans}\ \emph {et~al.}(2020)\citenamefont
  {Borjans}, \citenamefont {Croot}, \citenamefont {Mi}, \citenamefont
  {Gullans},\ and\ \citenamefont {Petta}}]{borjans2020resonant}%
  \BibitemOpen
  \bibfield  {author} {\bibinfo {author} {\bibfnamefont {F.}~\bibnamefont
  {Borjans}}, \bibinfo {author} {\bibfnamefont {X.}~\bibnamefont {Croot}},
  \bibinfo {author} {\bibfnamefont {X.}~\bibnamefont {Mi}}, \bibinfo {author}
  {\bibfnamefont {M.}~\bibnamefont {Gullans}}, \ and\ \bibinfo {author}
  {\bibfnamefont {J.}~\bibnamefont {Petta}},\ }\bibfield  {title} {\enquote
  {\bibinfo {title} {Resonant microwave-mediated interactions between distant
  electron spins},}\ }\href@noop {} {\bibfield  {journal} {\bibinfo  {journal}
  {Nature}\ }\textbf {\bibinfo {volume} {577}},\ \bibinfo {pages} {195}
  (\bibinfo {year} {2020})}\BibitemShut {NoStop}%
\bibitem [{\citenamefont {Hao}\ \emph {et~al.}(2010)\citenamefont {Hao},
  \citenamefont {Tu}, \citenamefont {Cao}, \citenamefont {Zhou}, \citenamefont
  {Li}, \citenamefont {Guo}, \citenamefont {Fung}, \citenamefont {Ji},
  \citenamefont {Guo},\ and\ \citenamefont {Lu}}]{doi:10.1021/nl101181e}%
  \BibitemOpen
  \bibfield  {author} {\bibinfo {author} {\bibfnamefont {X.-J.}\ \bibnamefont
  {Hao}}, \bibinfo {author} {\bibfnamefont {T.}~\bibnamefont {Tu}}, \bibinfo
  {author} {\bibfnamefont {G.}~\bibnamefont {Cao}}, \bibinfo {author}
  {\bibfnamefont {C.}~\bibnamefont {Zhou}}, \bibinfo {author} {\bibfnamefont
  {H.-O.}\ \bibnamefont {Li}}, \bibinfo {author} {\bibfnamefont {G.-C.}\
  \bibnamefont {Guo}}, \bibinfo {author} {\bibfnamefont {W.~Y.}\ \bibnamefont
  {Fung}}, \bibinfo {author} {\bibfnamefont {Z.}~\bibnamefont {Ji}}, \bibinfo
  {author} {\bibfnamefont {G.-P.}\ \bibnamefont {Guo}}, \ and\ \bibinfo
  {author} {\bibfnamefont {W.}~\bibnamefont {Lu}},\ }\bibfield  {title}
  {\enquote {\bibinfo {title} {Strong and Tunable Spin-Orbit Coupling of
  One-Dimensional Holes in Ge/Si Core/Shell Nanowires},}\ }\href {\doibase
  10.1021/nl101181e} {\bibfield  {journal} {\bibinfo  {journal} {Nano Lett.}\
  }\textbf {\bibinfo {volume} {10}},\ \bibinfo {pages} {2956} (\bibinfo {year}
  {2010})}\BibitemShut {NoStop}%
\bibitem [{\citenamefont {Kloeffel}\ \emph {et~al.}(2011)\citenamefont
  {Kloeffel}, \citenamefont {Trif},\ and\ \citenamefont
  {Loss}}]{PhysRevB.84.195314}%
  \BibitemOpen
  \bibfield  {author} {\bibinfo {author} {\bibfnamefont {C.}~\bibnamefont
  {Kloeffel}}, \bibinfo {author} {\bibfnamefont {M.}~\bibnamefont {Trif}}, \
  and\ \bibinfo {author} {\bibfnamefont {D.}~\bibnamefont {Loss}},\ }\bibfield
  {title} {\enquote {\bibinfo {title} {Strong spin-orbit interaction and
  helical hole states in Ge/Si nanowires},}\ }\href {\doibase
  10.1103/PhysRevB.84.195314} {\bibfield  {journal} {\bibinfo  {journal} {Phys.
  Rev. B}\ }\textbf {\bibinfo {volume} {84}},\ \bibinfo {pages} {195314}
  (\bibinfo {year} {2011})}\BibitemShut {NoStop}%
\bibitem [{\citenamefont {Hendrickx}\ \emph {et~al.}(2020)\citenamefont
  {Hendrickx}, \citenamefont {Franke}, \citenamefont {Sammak}, \citenamefont
  {Scappucci},\ and\ \citenamefont {Veldhorst}}]{hendrickx2020fast}%
  \BibitemOpen
  \bibfield  {author} {\bibinfo {author} {\bibfnamefont {N.}~\bibnamefont
  {Hendrickx}}, \bibinfo {author} {\bibfnamefont {D.}~\bibnamefont {Franke}},
  \bibinfo {author} {\bibfnamefont {A.}~\bibnamefont {Sammak}}, \bibinfo
  {author} {\bibfnamefont {G.}~\bibnamefont {Scappucci}}, \ and\ \bibinfo
  {author} {\bibfnamefont {M.}~\bibnamefont {Veldhorst}},\ }\bibfield  {title}
  {\enquote {\bibinfo {title} {Fast two-qubit logic with holes in germanium},}\
  }\href@noop {} {\bibfield  {journal} {\bibinfo  {journal} {Nature}\ }\textbf
  {\bibinfo {volume} {577}},\ \bibinfo {pages} {487} (\bibinfo {year}
  {2020})}\BibitemShut {NoStop}%
\bibitem [{\citenamefont {Froning}\ \emph {et~al.}(2021)\citenamefont
  {Froning}, \citenamefont {Camenzind}, \citenamefont {van~der Molen},
  \citenamefont {Li}, \citenamefont {Bakkers}, \citenamefont {Zumbühl},\ and\
  \citenamefont {Braakman}}]{Ultrafast2021}%
  \BibitemOpen
  \bibfield  {author} {\bibinfo {author} {\bibfnamefont {F.~N.~M.}\
  \bibnamefont {Froning}}, \bibinfo {author} {\bibfnamefont {L.~C.}\
  \bibnamefont {Camenzind}}, \bibinfo {author} {\bibfnamefont {O.~A.~H.}\
  \bibnamefont {van~der Molen}}, \bibinfo {author} {\bibfnamefont
  {A.}~\bibnamefont {Li}}, \bibinfo {author} {\bibfnamefont {E.~P. A.~M.}\
  \bibnamefont {Bakkers}}, \bibinfo {author} {\bibfnamefont {D.~M.}\
  \bibnamefont {Zumbühl}}, \ and\ \bibinfo {author} {\bibfnamefont {F.~R.}\
  \bibnamefont {Braakman}},\ }\bibfield  {title} {\enquote {\bibinfo {title}
  {Ultrafast hole spin qubit with gate-tunable spin–orbit switch
  functionality},}\ }\href {https://doi.org/10.1038/s41565-020-00828-6}
  {\bibfield  {journal} {\bibinfo  {journal} {Nat. Nanotechnol.}\ }\textbf
  {\bibinfo {volume} {16}},\ \bibinfo {pages} {308} (\bibinfo {year}
  {2021})}\BibitemShut {NoStop}%
\bibitem [{\citenamefont {Hsu}\ \emph {et~al.}(2016)\citenamefont {Hsu},
  \citenamefont {Zhen}, \citenamefont {Stone}, \citenamefont {Joannopoulos},\
  and\ \citenamefont {Solja{\v c}i{\'c}}}]{Hsu2016}%
  \BibitemOpen
  \bibfield  {author} {\bibinfo {author} {\bibfnamefont {C.~W.}\ \bibnamefont
  {Hsu}}, \bibinfo {author} {\bibfnamefont {B.}~\bibnamefont {Zhen}}, \bibinfo
  {author} {\bibfnamefont {A.~D.}\ \bibnamefont {Stone}}, \bibinfo {author}
  {\bibfnamefont {J.~D.}\ \bibnamefont {Joannopoulos}}, \ and\ \bibinfo
  {author} {\bibfnamefont {M.}~\bibnamefont {Solja{\v c}i{\'c}}},\ }\bibfield
  {title} {\enquote {\bibinfo {title} {Bound states in the continuum},}\ }\href
  {https://doi.org/10.1038/natrevmats.2016.48} {\bibfield  {journal} {\bibinfo
  {journal} {Nat. Rev. Mater.}\ }\textbf {\bibinfo {volume} {1}},\ \bibinfo
  {pages} {16048} (\bibinfo {year} {2016})}\BibitemShut {NoStop}%
\bibitem [{\citenamefont {Overvig}\ \emph {et~al.}(2021)\citenamefont
  {Overvig}, \citenamefont {Yu},\ and\ \citenamefont
  {Al{\`u}}}]{PhysRevLett.126.073001}%
  \BibitemOpen
  \bibfield  {author} {\bibinfo {author} {\bibfnamefont {A.}~\bibnamefont
  {Overvig}}, \bibinfo {author} {\bibfnamefont {N.}~\bibnamefont {Yu}}, \ and\
  \bibinfo {author} {\bibfnamefont {A.}~\bibnamefont {Al{\`u}}},\ }\bibfield
  {title} {\enquote {\bibinfo {title} {Chiral Quasi-Bound States in the
  Continuum},}\ }\href {\doibase 10.1103/PhysRevLett.126.073001} {\bibfield
  {journal} {\bibinfo  {journal} {Phys. Rev. Lett.}\ }\textbf {\bibinfo
  {volume} {126}},\ \bibinfo {pages} {073001} (\bibinfo {year}
  {2021})}\BibitemShut {NoStop}%
\bibitem [{\citenamefont {Vallejo}\ \emph {et~al.}(2010)\citenamefont
  {Vallejo}, \citenamefont {{Ladr{\'o}n de Guevara}},\ and\ \citenamefont
  {Orellana}}]{VALLEJO20104928}%
  \BibitemOpen
  \bibfield  {author} {\bibinfo {author} {\bibfnamefont {M.}~\bibnamefont
  {Vallejo}}, \bibinfo {author} {\bibfnamefont {M.}~\bibnamefont {{Ladr{\'o}n
  de Guevara}}}, \ and\ \bibinfo {author} {\bibfnamefont {P.}~\bibnamefont
  {Orellana}},\ }\bibfield  {title} {\enquote {\bibinfo {title} {Triple Rashba
  dots as a spin filter: Bound states in the continuum and Fano effect},}\
  }\href {\doibase 10.1016/j.physleta.2010.10.015} {\bibfield  {journal}
  {\bibinfo  {journal} {Phys. Lett. A}\ }\textbf {\bibinfo {volume} {374}},\
  \bibinfo {pages} {4928} (\bibinfo {year} {2010})}\BibitemShut {NoStop}%
\bibitem [{\citenamefont {Lin}\ \emph {et~al.}(2013)\citenamefont {Lin},
  \citenamefont {Mueller}, \citenamefont {Wang}, \citenamefont {Yuan},
  \citenamefont {Antoniou}, \citenamefont {Yuan},\ and\ \citenamefont
  {Capasso}}]{Lin331}%
  \BibitemOpen
  \bibfield  {author} {\bibinfo {author} {\bibfnamefont {J.}~\bibnamefont
  {Lin}}, \bibinfo {author} {\bibfnamefont {J.~P.~B.}\ \bibnamefont {Mueller}},
  \bibinfo {author} {\bibfnamefont {Q.}~\bibnamefont {Wang}}, \bibinfo {author}
  {\bibfnamefont {G.}~\bibnamefont {Yuan}}, \bibinfo {author} {\bibfnamefont
  {N.}~\bibnamefont {Antoniou}}, \bibinfo {author} {\bibfnamefont {X.-C.}\
  \bibnamefont {Yuan}}, \ and\ \bibinfo {author} {\bibfnamefont
  {F.}~\bibnamefont {Capasso}},\ }\bibfield  {title} {\enquote {\bibinfo
  {title} {Polarization-Controlled Tunable Directional Coupling of Surface
  Plasmon Polaritons},}\ }\href {\doibase 10.1126/science.1233746} {\bibfield
  {journal} {\bibinfo  {journal} {Science}\ }\textbf {\bibinfo {volume}
  {340}},\ \bibinfo {pages} {331} (\bibinfo {year} {2013})}\BibitemShut
  {NoStop}%
\bibitem [{\citenamefont {Spitzer}\ \emph {et~al.}(2018)\citenamefont
  {Spitzer}, \citenamefont {Poddubny}, \citenamefont {Akimov}, \citenamefont
  {Sapega}, \citenamefont {Klompmaker}, \citenamefont {Kreilkamp},
  \citenamefont {Litvin}, \citenamefont {Jede}, \citenamefont {Karczewski},
  \citenamefont {Wiater}, \citenamefont {Wojtowicz}, \citenamefont {Yakovlev},\
  and\ \citenamefont {Bayer}}]{Spitzer2018}%
  \BibitemOpen
  \bibfield  {author} {\bibinfo {author} {\bibfnamefont {F.}~\bibnamefont
  {Spitzer}}, \bibinfo {author} {\bibfnamefont {A.~N.}\ \bibnamefont
  {Poddubny}}, \bibinfo {author} {\bibfnamefont {I.~A.}\ \bibnamefont
  {Akimov}}, \bibinfo {author} {\bibfnamefont {V.~F.}\ \bibnamefont {Sapega}},
  \bibinfo {author} {\bibfnamefont {L.}~\bibnamefont {Klompmaker}}, \bibinfo
  {author} {\bibfnamefont {L.~E.}\ \bibnamefont {Kreilkamp}}, \bibinfo {author}
  {\bibfnamefont {L.~V.}\ \bibnamefont {Litvin}}, \bibinfo {author}
  {\bibfnamefont {R.}~\bibnamefont {Jede}}, \bibinfo {author} {\bibfnamefont
  {G.}~\bibnamefont {Karczewski}}, \bibinfo {author} {\bibfnamefont
  {M.}~\bibnamefont {Wiater}}, \bibinfo {author} {\bibfnamefont
  {T.}~\bibnamefont {Wojtowicz}}, \bibinfo {author} {\bibfnamefont {D.~R.}\
  \bibnamefont {Yakovlev}}, \ and\ \bibinfo {author} {\bibfnamefont
  {M.}~\bibnamefont {Bayer}},\ }\bibfield  {title} {\enquote {\bibinfo {title}
  {Routing the emission of a near-surface light source by a magnetic field},}\
  }\href {https://doi.org/10.1038/s41567-018-0232-7} {\bibfield  {journal}
  {\bibinfo  {journal} {Nat. Phys.}\ }\textbf {\bibinfo {volume} {14}},\
  \bibinfo {pages} {1043} (\bibinfo {year} {2018})}\BibitemShut {NoStop}%
\bibitem [{\citenamefont {Yin}\ \emph {et~al.}(2020)\citenamefont {Yin},
  \citenamefont {Jin}, \citenamefont {Solja{\v c}i{\'c}}, \citenamefont
  {Peng},\ and\ \citenamefont {Zhen}}]{Yin2020}%
  \BibitemOpen
  \bibfield  {author} {\bibinfo {author} {\bibfnamefont {X.}~\bibnamefont
  {Yin}}, \bibinfo {author} {\bibfnamefont {J.}~\bibnamefont {Jin}}, \bibinfo
  {author} {\bibfnamefont {M.}~\bibnamefont {Solja{\v c}i{\'c}}}, \bibinfo
  {author} {\bibfnamefont {C.}~\bibnamefont {Peng}}, \ and\ \bibinfo {author}
  {\bibfnamefont {B.}~\bibnamefont {Zhen}},\ }\bibfield  {title} {\enquote
  {\bibinfo {title} {Observation of topologically enabled unidirectional guided
  resonances},}\ }\href {https://doi.org/10.1038/s41586-020-2181-4} {\bibfield
  {journal} {\bibinfo  {journal} {Nature}\ }\textbf {\bibinfo {volume} {580}},\
  \bibinfo {pages} {467} (\bibinfo {year} {2020})}\BibitemShut {NoStop}%
\bibitem [{\citenamefont {Sato}\ \emph {et~al.}(2005)\citenamefont {Sato},
  \citenamefont {Aikawa}, \citenamefont {Kobayashi}, \citenamefont
  {Katsumoto},\ and\ \citenamefont {Iye}}]{PhysRevLett.95.066801}%
  \BibitemOpen
  \bibfield  {author} {\bibinfo {author} {\bibfnamefont {M.}~\bibnamefont
  {Sato}}, \bibinfo {author} {\bibfnamefont {H.}~\bibnamefont {Aikawa}},
  \bibinfo {author} {\bibfnamefont {K.}~\bibnamefont {Kobayashi}}, \bibinfo
  {author} {\bibfnamefont {S.}~\bibnamefont {Katsumoto}}, \ and\ \bibinfo
  {author} {\bibfnamefont {Y.}~\bibnamefont {Iye}},\ }\bibfield  {title}
  {\enquote {\bibinfo {title} {Observation of the Fano-Kondo Antiresonance in a
  Quantum Wire with a Side-Coupled Quantum Dot},}\ }\href {\doibase
  10.1103/PhysRevLett.95.066801} {\bibfield  {journal} {\bibinfo  {journal}
  {Phys. Rev. Lett.}\ }\textbf {\bibinfo {volume} {95}},\ \bibinfo {pages}
  {066801} (\bibinfo {year} {2005})}\BibitemShut {NoStop}%
\bibitem [{\citenamefont {Edlbauer}\ \emph {et~al.}(2017)\citenamefont
  {Edlbauer}, \citenamefont {Takada}, \citenamefont {Roussely}, \citenamefont
  {Yamamoto}, \citenamefont {Tarucha}, \citenamefont {Ludwig}, \citenamefont
  {Wieck}, \citenamefont {Meunier},\ and\ \citenamefont
  {B{\"a}uerle}}]{Edlbauer2017}%
  \BibitemOpen
  \bibfield  {author} {\bibinfo {author} {\bibfnamefont {H.}~\bibnamefont
  {Edlbauer}}, \bibinfo {author} {\bibfnamefont {S.}~\bibnamefont {Takada}},
  \bibinfo {author} {\bibfnamefont {G.}~\bibnamefont {Roussely}}, \bibinfo
  {author} {\bibfnamefont {M.}~\bibnamefont {Yamamoto}}, \bibinfo {author}
  {\bibfnamefont {S.}~\bibnamefont {Tarucha}}, \bibinfo {author} {\bibfnamefont
  {A.}~\bibnamefont {Ludwig}}, \bibinfo {author} {\bibfnamefont {A.~D.}\
  \bibnamefont {Wieck}}, \bibinfo {author} {\bibfnamefont {T.}~\bibnamefont
  {Meunier}}, \ and\ \bibinfo {author} {\bibfnamefont {C.}~\bibnamefont
  {B{\"a}uerle}},\ }\bibfield  {title} {\enquote {\bibinfo {title}
  {Non-universal transmission phase behaviour of a large quantum dot},}\ }\href
  {https://doi.org/10.1038/s41467-017-01685-z} {\bibfield  {journal} {\bibinfo
  {journal} {Nat. Commun.}\ }\textbf {\bibinfo {volume} {8}},\ \bibinfo {pages}
  {1710} (\bibinfo {year} {2017})}\BibitemShut {NoStop}%
\bibitem [{\citenamefont {{V. Borzenets}}\ \emph {et~al.}(2020)\citenamefont
  {{V. Borzenets}}, \citenamefont {Shim}, \citenamefont {Chen}, \citenamefont
  {Ludwig}, \citenamefont {Wieck}, \citenamefont {Tarucha}, \citenamefont
  {Sim},\ and\ \citenamefont {Yamamoto}}]{Borzenets2020}%
  \BibitemOpen
  \bibfield  {author} {\bibinfo {author} {\bibfnamefont {I.}~\bibnamefont {{V.
  Borzenets}}}, \bibinfo {author} {\bibfnamefont {J.}~\bibnamefont {Shim}},
  \bibinfo {author} {\bibfnamefont {J.~C.~H.}\ \bibnamefont {Chen}}, \bibinfo
  {author} {\bibfnamefont {A.}~\bibnamefont {Ludwig}}, \bibinfo {author}
  {\bibfnamefont {A.~D.}\ \bibnamefont {Wieck}}, \bibinfo {author}
  {\bibfnamefont {S.}~\bibnamefont {Tarucha}}, \bibinfo {author} {\bibfnamefont
  {H.-S.}\ \bibnamefont {Sim}}, \ and\ \bibinfo {author} {\bibfnamefont
  {M.}~\bibnamefont {Yamamoto}},\ }\bibfield  {title} {\enquote {\bibinfo
  {title} {Observation of the Kondo screening cloud},}\ }\href
  {https://doi.org/10.1038/s41586-020-2058-6} {\bibfield  {journal} {\bibinfo
  {journal} {Nature}\ }\textbf {\bibinfo {volume} {579}},\ \bibinfo {pages}
  {210} (\bibinfo {year} {2020})}\BibitemShut {NoStop}%
\bibitem [{\citenamefont {Luttinger}(1956)}]{PhysRev.102.1030}%
  \BibitemOpen
  \bibfield  {author} {\bibinfo {author} {\bibfnamefont {J.~M.}\ \bibnamefont
  {Luttinger}},\ }\bibfield  {title} {\enquote {\bibinfo {title} {Quantum
  Theory of Cyclotron Resonance in Semiconductors: General Theory},}\ }\href
  {\doibase 10.1103/PhysRev.102.1030} {\bibfield  {journal} {\bibinfo
  {journal} {Phys. Rev.}\ }\textbf {\bibinfo {volume} {102}},\ \bibinfo {pages}
  {1030} (\bibinfo {year} {1956})}\BibitemShut {NoStop}%
\bibitem [{\citenamefont {Ivchenko}(2005)}]{ivchenko05a}%
  \BibitemOpen
  \bibfield  {author} {\bibinfo {author} {\bibfnamefont {E.~L.}\ \bibnamefont
  {Ivchenko}},\ }\href@noop {} {\emph {\bibinfo {title} {Optical spectroscopy
  of semiconductor nanostructures}}}\ (\bibinfo  {publisher} {Alpha Science,
  Harrow UK},\ \bibinfo {year} {2005})\BibitemShut {NoStop}%
\bibitem [{sup()}]{supp}%
  \BibitemOpen
  \href@noop {} {\enquote {\bibinfo {title} {See Supplemental Material for the detailed calculation of the tunneling matrix elements and the Keldysh formalism for arbitrary interaction strength, as well as for the detailed analysis of the limits of weak and strong interaction}}\ }\BibitemShut {NoStop}%
\bibitem [{\citenamefont {Anderson}(1961)}]{PhysRev.124.41}%
  \BibitemOpen
  \bibfield  {author} {\bibinfo {author} {\bibfnamefont {P.~W.}\ \bibnamefont
  {Anderson}},\ }\bibfield  {title} {\enquote {\bibinfo {title} {Localized
  Magnetic States in Metals},}\ }\href {\doibase 10.1103/PhysRev.124.41}
  {\bibfield  {journal} {\bibinfo  {journal} {Phys. Rev.}\ }\textbf {\bibinfo
  {volume} {124}},\ \bibinfo {pages} {41} (\bibinfo {year} {1961})}\BibitemShut
  {NoStop}%
\bibitem [{\citenamefont {Fano}(1961)}]{fano61}%
  \BibitemOpen
  \bibfield  {author} {\bibinfo {author} {\bibfnamefont {U.}~\bibnamefont
  {Fano}},\ }\bibfield  {title} {\enquote {\bibinfo {title} {Effects of
  Configuration Interaction on Intensities and Phase Shifts},}\ }\href@noop {}
  {\bibfield  {journal} {\bibinfo  {journal} {Phys. Rev.}\ }\textbf {\bibinfo
  {volume} {124}},\ \bibinfo {pages} {1866} (\bibinfo {year}
  {1961})}\BibitemShut {NoStop}%
\bibitem [{\citenamefont {Limonov}\ \emph {et~al.}(2017)\citenamefont
  {Limonov}, \citenamefont {Rybin}, \citenamefont {Poddubny},\ and\
  \citenamefont {Kivshar}}]{Poddubny_Fano}%
  \BibitemOpen
  \bibfield  {author} {\bibinfo {author} {\bibfnamefont {M.~F.}\ \bibnamefont
  {Limonov}}, \bibinfo {author} {\bibfnamefont {M.~V.}\ \bibnamefont {Rybin}},
  \bibinfo {author} {\bibfnamefont {A.~N.}\ \bibnamefont {Poddubny}}, \ and\
  \bibinfo {author} {\bibfnamefont {Y.~S.}\ \bibnamefont {Kivshar}},\
  }\bibfield  {title} {\enquote {\bibinfo {title} {Fano resonances in
  photonics},}\ }\href {https://doi.org/10.1038/nphoton.2017.142} {\bibfield
  {journal} {\bibinfo  {journal} {Nat. Photon.}\ }\textbf {\bibinfo {volume}
  {11}},\ \bibinfo {pages} {543} (\bibinfo {year} {2017})}\BibitemShut
  {NoStop}%
\bibitem [{\citenamefont {Lodahl}\ \emph {et~al.}(2017)\citenamefont {Lodahl},
  \citenamefont {Mahmoodian}, \citenamefont {Stobbe}, \citenamefont
  {Rauschenbeutel}, \citenamefont {Schneeweiss}, \citenamefont {Volz},
  \citenamefont {Pichler},\ and\ \citenamefont {Zoller}}]{lodahl2017chiral}%
  \BibitemOpen
  \bibfield  {author} {\bibinfo {author} {\bibfnamefont {P.}~\bibnamefont
  {Lodahl}}, \bibinfo {author} {\bibfnamefont {S.}~\bibnamefont {Mahmoodian}},
  \bibinfo {author} {\bibfnamefont {S.}~\bibnamefont {Stobbe}}, \bibinfo
  {author} {\bibfnamefont {A.}~\bibnamefont {Rauschenbeutel}}, \bibinfo
  {author} {\bibfnamefont {P.}~\bibnamefont {Schneeweiss}}, \bibinfo {author}
  {\bibfnamefont {J.}~\bibnamefont {Volz}}, \bibinfo {author} {\bibfnamefont
  {H.}~\bibnamefont {Pichler}}, \ and\ \bibinfo {author} {\bibfnamefont
  {P.}~\bibnamefont {Zoller}},\ }\bibfield  {title} {\enquote {\bibinfo {title}
  {Chiral quantum optics},}\ }\href {https://doi.org/10.1038/nature21037}
  {\bibfield  {journal} {\bibinfo  {journal} {Nature}\ }\textbf {\bibinfo
  {volume} {541}},\ \bibinfo {pages} {473} (\bibinfo {year}
  {2017})}\BibitemShut {NoStop}%
\bibitem [{\citenamefont {Rammer}\ and\ \citenamefont
  {Smith}(1986)}]{RevModPhys.58.323}%
  \BibitemOpen
  \bibfield  {author} {\bibinfo {author} {\bibfnamefont {J.}~\bibnamefont
  {Rammer}}\ and\ \bibinfo {author} {\bibfnamefont {H.}~\bibnamefont {Smith}},\
  }\bibfield  {title} {\enquote {\bibinfo {title} {Quantum field-theoretical
  methods in transport theory of metals},}\ }\href {\doibase
  10.1103/RevModPhys.58.323} {\bibfield  {journal} {\bibinfo  {journal} {Rev.
  Mod. Phys.}\ }\textbf {\bibinfo {volume} {58}},\ \bibinfo {pages} {323}
  (\bibinfo {year} {1986})}\BibitemShut {NoStop}%
\bibitem [{\citenamefont {Stefanucci}\ and\ \citenamefont {van
  Leeuwen}(2013)}]{stefanucci_vanleeuwen_2013}%
  \BibitemOpen
  \bibfield  {author} {\bibinfo {author} {\bibfnamefont {G.}~\bibnamefont
  {Stefanucci}}\ and\ \bibinfo {author} {\bibfnamefont {R.}~\bibnamefont {van
  Leeuwen}},\ }\href {\doibase 10.1017/CBO9781139023979} {\emph {\bibinfo
  {title} {Nonequilibrium Many-Body Theory of Quantum Systems: A Modern
  Introduction}}}\ (\bibinfo  {publisher} {Cambridge University Press},\
  \bibinfo {year} {2013})\BibitemShut {NoStop}%
\bibitem [{\citenamefont {Arseev}(2015)}]{Arseev_2015}%
  \BibitemOpen
  \bibfield  {author} {\bibinfo {author} {\bibfnamefont {P.~I.}\ \bibnamefont
  {Arseev}},\ }\bibfield  {title} {\enquote {\bibinfo {title} {On the
  nonequilibrium diagram technique: derivation, some features, and
  applications},}\ }\href {\doibase 10.3367/ufne.0185.201512b.1271} {\bibfield
  {journal} {\bibinfo  {journal} {Phys. Usp.}\ }\textbf {\bibinfo {volume}
  {58}},\ \bibinfo {pages} {1159} (\bibinfo {year} {2015})}\BibitemShut
  {NoStop}%
\bibitem [{\citenamefont {Hubbard}\ and\ \citenamefont
  {Flowers}(1963)}]{doi:10.1098/rspa.1963.0204}%
  \BibitemOpen
  \bibfield  {author} {\bibinfo {author} {\bibfnamefont {J.}~\bibnamefont
  {Hubbard}}\ and\ \bibinfo {author} {\bibfnamefont {B.~H.}\ \bibnamefont
  {Flowers}},\ }\bibfield  {title} {\enquote {\bibinfo {title} {Electron
  correlations in narrow energy bands},}\ }\href {\doibase
  10.1098/rspa.1963.0204} {\bibfield  {journal} {\bibinfo  {journal} {J. Proc.
  Roy. Soc. A}\ }\textbf {\bibinfo {volume} {276}},\ \bibinfo {pages} {238}
  (\bibinfo {year} {1963})}\BibitemShut {NoStop}%
\bibitem [{\citenamefont {Haug}\ and\ \citenamefont
  {Jauho}(2008)}]{haug2008quantum}%
  \BibitemOpen
  \bibfield  {author} {\bibinfo {author} {\bibfnamefont {H.}~\bibnamefont
  {Haug}}\ and\ \bibinfo {author} {\bibfnamefont {A.-P.}\ \bibnamefont
  {Jauho}},\ }\href@noop {} {\emph {\bibinfo {title} {Quantum kinetics in
  transport and optics of semiconductors}}},\ Vol.~\bibinfo {volume} {2}\
  (\bibinfo  {publisher} {Springer},\ \bibinfo {year} {2008})\BibitemShut
  {NoStop}%
\bibitem [{\citenamefont {Lacroix}(1981)}]{Lacroix_1981}%
  \BibitemOpen
  \bibfield  {author} {\bibinfo {author} {\bibfnamefont {C.}~\bibnamefont
  {Lacroix}},\ }\bibfield  {title} {\enquote {\bibinfo {title} {Density of
  states for the Anderson model},}\ }\href {\doibase
  10.1088/0305-4608/11/11/020} {\bibfield  {journal} {\bibinfo  {journal} {J.
  Phys. F}\ }\textbf {\bibinfo {volume} {11}},\ \bibinfo {pages} {2389}
  (\bibinfo {year} {1981})}\BibitemShut {NoStop}%
\bibitem [{\citenamefont {Meir}\ \emph {et~al.}(1993)\citenamefont {Meir},
  \citenamefont {Wingreen},\ and\ \citenamefont {Lee}}]{PhysRevLett.70.2601}%
  \BibitemOpen
  \bibfield  {author} {\bibinfo {author} {\bibfnamefont {Y.}~\bibnamefont
  {Meir}}, \bibinfo {author} {\bibfnamefont {N.~S.}\ \bibnamefont {Wingreen}},
  \ and\ \bibinfo {author} {\bibfnamefont {P.~A.}\ \bibnamefont {Lee}},\
  }\bibfield  {title} {\enquote {\bibinfo {title} {Low-temperature transport
  through a quantum dot: The Anderson model out of equilibrium},}\ }\href
  {\doibase 10.1103/PhysRevLett.70.2601} {\bibfield  {journal} {\bibinfo
  {journal} {Phys. Rev. Lett.}\ }\textbf {\bibinfo {volume} {70}},\ \bibinfo
  {pages} {2601} (\bibinfo {year} {1993})}\BibitemShut {NoStop}%
\bibitem [{\citenamefont {\ifmmode~\acute{S}\else \'{S}\fi{}wirkowicz}\ \emph
  {et~al.}(2003)\citenamefont {\ifmmode~\acute{S}\else \'{S}\fi{}wirkowicz},
  \citenamefont {Barna\ifmmode~\acute{s}\else \'{s}\fi{}},\ and\ \citenamefont
  {Wilczy\ifmmode~\acute{n}\else \'{n}\fi{}ski}}]{PhysRevB.68.195318}%
  \BibitemOpen
  \bibfield  {author} {\bibinfo {author} {\bibfnamefont {R.}~\bibnamefont
  {\ifmmode~\acute{S}\else \'{S}\fi{}wirkowicz}}, \bibinfo {author}
  {\bibfnamefont {J.}~\bibnamefont {Barna\ifmmode~\acute{s}\else \'{s}\fi{}}},
  \ and\ \bibinfo {author} {\bibfnamefont {M.}~\bibnamefont
  {Wilczy\ifmmode~\acute{n}\else \'{n}\fi{}ski}},\ }\bibfield  {title}
  {\enquote {\bibinfo {title} {Nonequilibrium Kondo effect in quantum dots},}\
  }\href {\doibase 10.1103/PhysRevB.68.195318} {\bibfield  {journal} {\bibinfo
  {journal} {Phys. Rev. B}\ }\textbf {\bibinfo {volume} {68}},\ \bibinfo
  {pages} {195318} (\bibinfo {year} {2003})}\BibitemShut {NoStop}%
\bibitem [{\citenamefont {Glazov}(2018)}]{book_Glazov}%
  \BibitemOpen
  \bibfield  {author} {\bibinfo {author} {\bibfnamefont {M.~M.}\ \bibnamefont
  {Glazov}},\ }\href@noop {} {\emph {\bibinfo {title} {Electron and Nuclear
  Spin Dynamics in Semiconductor Nanostructures}}}\ (\bibinfo  {publisher}
  {Oxford University Press, Oxford},\ \bibinfo {year} {2018})\BibitemShut
  {NoStop}%
\bibitem [{\citenamefont {Smirnov}\ \emph {et~al.}(2020)\citenamefont
  {Smirnov}, \citenamefont {Zhukov}, \citenamefont {Yakovlev}, \citenamefont
  {Kirstein}, \citenamefont {Bayer},\ and\ \citenamefont {Greilich}}]{PRC}%
  \BibitemOpen
  \bibfield  {author} {\bibinfo {author} {\bibfnamefont {D.~S.}\ \bibnamefont
  {Smirnov}}, \bibinfo {author} {\bibfnamefont {E.~A.}\ \bibnamefont {Zhukov}},
  \bibinfo {author} {\bibfnamefont {D.~R.}\ \bibnamefont {Yakovlev}}, \bibinfo
  {author} {\bibfnamefont {E.}~\bibnamefont {Kirstein}}, \bibinfo {author}
  {\bibfnamefont {M.}~\bibnamefont {Bayer}}, \ and\ \bibinfo {author}
  {\bibfnamefont {A.}~\bibnamefont {Greilich}},\ }\bibfield  {title} {\enquote
  {\bibinfo {title} {Spin polarization recovery and Hanle effect for charge
  carriers interacting with nuclear spins in semiconductors},}\ }\href
  {\doibase 10.1103/PhysRevB.102.235413} {\bibfield  {journal} {\bibinfo
  {journal} {Phys. Rev. B}\ }\textbf {\bibinfo {volume} {102}},\ \bibinfo
  {pages} {235413} (\bibinfo {year} {2020})}\BibitemShut {NoStop}%
\bibitem [{\citenamefont {Hanson}\ \emph {et~al.}(2007)\citenamefont {Hanson},
  \citenamefont {Kouwenhoven}, \citenamefont {Petta}, \citenamefont {Tarucha},\
  and\ \citenamefont {Vandersypen}}]{hanson07}%
  \BibitemOpen
  \bibfield  {author} {\bibinfo {author} {\bibfnamefont {R.}~\bibnamefont
  {Hanson}}, \bibinfo {author} {\bibfnamefont {L.~P.}\ \bibnamefont
  {Kouwenhoven}}, \bibinfo {author} {\bibfnamefont {J.~R.}\ \bibnamefont
  {Petta}}, \bibinfo {author} {\bibfnamefont {S.}~\bibnamefont {Tarucha}}, \
  and\ \bibinfo {author} {\bibfnamefont {L.~M.~K.}\ \bibnamefont
  {Vandersypen}},\ }\bibfield  {title} {\enquote {\bibinfo {title} {Spins in
  few-electron quantum dots},}\ }\href@noop {} {\bibfield  {journal} {\bibinfo
  {journal} {Rev. Mod. Phys}\ }\textbf {\bibinfo {volume} {79}},\ \bibinfo
  {pages} {1217} (\bibinfo {year} {2007})}\BibitemShut {NoStop}%
\end{thebibliography}

\end{document}